\documentclass[11pt,a4paper]{article}
\usepackage[utf8]{inputenc}
\usepackage{epic,eepic,epsfig,amsmath,amssymb,amsthm,mathabx}
\usepackage{t1enc}

\usepackage[francais,english]{babel}

\usepackage{caption}

\usepackage{color}

\usepackage{bbm}

\def\virgp{\raise 2pt\hbox{,}}

\renewcommand{\geq}{\geqslant}

\def\N{{\mathbb N}}

\def\R{{\mathbb R}}

\def\virgp{\raise 2pt\hbox{,}}
\def\cdotpv{\raise 2pt\hbox{;}}

\def\1{\mathbbm{1}}

\newtheorem{theorem}{Theorem}[section]

\newtheorem{proposition}[theorem]{Proposition}
\newtheorem{lemma}[theorem]{Lemma}

\newtheorem{pte}[theorem]{Property}

\theoremstyle{remark}
\newtheorem{remark}{Remark}[section]

\theoremstyle{definition}
\newtheorem{definition}{Definition}[section]

\newtheorem*{notation}{Notation}

\newtheorem*{notations}{Notations}

\theoremstyle{definition}

\theoremstyle{definition}

\newtheorem{assumptions}[theorem]{Assumptions}

\begin{document}
	
	\title{Generalized measure Black-Scholes equation: Towards option self-similar pricing}
	
	\author{Nizar Riane${^\dag}{^\ddag}$, Claire David$^\ddag$}

	\maketitle
	\centerline{$^\dag$ Universit\'e Mohammed V de Rabat, FSJESR-Agdal, Maroc\footnote{nizar.riane@gmail.com} }
	
	\vskip 0.5cm
	
	\centerline{$^\ddag$ Sorbonne Universit\'e}
	
	\centerline{CNRS, UMR 7598, Laboratoire Jacques-Louis Lions, 4, place Jussieu 75005, Paris, France\footnote{Corresponding author: Claire.David@Sorbonne-Universite.fr}}

	\begin{abstract}
		In this work, we give a generalized formulation of the Black-Scholes model. The novelty resides in considering the Black-Scholes model to be valid on 'average', but such that the pointwise option price dynamics depends on a measure representing the investors' 'uncertainty'.
		
		We make use of the theory of non-symmetric Dirichlet forms and the abstract theory of partial differential equations to establish well posedness of the problem. A detailed numerical analysis is given in the case of self-similar measures.
	\end{abstract}

	\maketitle
	\vskip 1cm
	
	
	\vskip 1cm
	
	
	\vskip 1cm
	
	\section{Introduction}
	
	\vskip 1cm
	
\hskip 0.5cm The Black-Scholes model arises as one of the most important application of mathematics to economics and finance of the 20$^{th}$ century, allowing the emergence of the theory of financial partial differential equations and the associated numerical analysis.\\
 
Yet, despite its phenomenal success in the financial market, the model suffers from deficiencies, for instance, when it comes to the modeling of real market options with erratic behaviors~\cite{Fortune1996}. One can argue that the model does not capture all the factors influencing investor decisions.\\

From a practical point of view, investors employ technical analysis to detect patterns in price evolution graphs to beat the market; since those patterns seem to appear in small and large scales, this could justify a self-similar modelling of the market. Investigation of fractals in finance is not new, we can refer, for instance, to Beno\^it Mandelbrot's work in~\cite{Mandelbrot1997}, the aspects of which are not taken into account by the classical model.\\
	
In this paper, we give a definition of a general version of the Black-Scholes model, based on the theory of non-symmetric Dirichlet forms, and on the abstract theory of partial differential equations. The key point is to consider the Black-Scholes equation describing an average evolution, while the exact dynamics depends on uncertainty captured by a mathematical measure. The analysis goes so far as proving the existence of a \emph{Generalized Black-Scholes operator}.\\
	
Special treatment is given to the self-similar case by writing an explicit formula, which enables computation of the solution. The CFL\footnote{Courant-Friedrichs-Lewy.} convergence condition for the associated finite difference scheme is determined and written explicitly. For the proper calibration, our model can avoid overpricing options \textbf{at the money}, and underpricing options at the ends, either deep \textbf{in the money}\footnote{When the option has what is also called an \emph{intrinsic value}, i.e. the real value of the option, that is to say the profit that could be made in the event of immediate exercise. It means that the value is at a favorable strike price relative to the prevailing market price of the underlying asset. Yet, this does not mean that the trader will be making profit, since the expense of buying, and the commission prices have also to be considered.}, or deep \textbf{out of the money}\footnote{When the option has what is also called an \emph{extrinsic value}, i.e. a value at a strike price higher than the market price of the underlying asset. In such a case, the Delta, i.e. the Greek which quantifies the risk, is less than~50.}. This work opens the way to an empirical investigation and an inverse problem of the probability measure $\mu$.\\
	
\section{Generalized Measure Black-Scholes Model}

\vskip 1cm

\begin{definition}[\textbf{Generalized Measure Black-Scholes Equation}] $\, $\\
	
	Let's introduce the generalized measure Black-Scholes equation, for European options, in the sense of distributions:
	
	$$\left \lbrace 
	\begin{array}{ccccc}
	\displaystyle \frac{\partial u}{\partial t}(t,x)  \, d\mu&=& 	\displaystyle \left(- r(t) \, x \,\frac{\partial u}{\partial x}(t,x) - \frac{\sigma^2(t)}{2}\,x^2 \,\Delta u(t,x)  + r(t) \, u(t,x) \right) \, dx &\quad \forall  \,(t,x) \, \in\,  \left[0,T\right] \times  \left]L,M\right[ \\
	u(T,x)&=&h(x)  &\quad \forall \, x \, \in \,  \left[L,M\right]
	\end{array}
	\right.$$
	
	\noindent where the variable $0<L<x<M$ is the price of the underlying financial instrument, $\sigma$ denotes the volatility,~$r$ the risk-free interest rate,~$T$ the maturity of the option, $\mu$ a non atomic finite measure with total mass $\mathbb{M}=M-L$ and~$u$ represents the option price. The real valued function $h$ takes the values $h(x)=(x-K)^+$ for a call\footnote{The \emph{call} is an option on a financial instrument, which consists in a right to buy. Concretely, it consists in a contract which allows the subscriber to get the targeted financial product, at a price fixed in advance - the \emph{strike} price - at a given date - the expiry one, or \emph{maturity} of the call.}, and $h(x)=(K-x)^+$ for a \emph{put}\footnote{As for the \emph{put}, it is this time a right to sell - or not - at the maturity date.}, given a constant~$K$: ``the strike".

	\end{definition}

	\vskip 1cm

	In order to prove that the problem has a solution, technical results associated to the classical Black-Scholes model are required. We refer to~\cite{Achdou2005} for the proof of the following assertions.

	\newpage

\subsection{The Black-Scholes Model}

\vskip 1cm

	\begin{notation}[\textbf{Space of Test Functions}] $\, $\\

		Given a continuous subset~$E$ of~$\R$, we will denote by~\mbox{${\cal D}(E)$} the space of test functions on~$E$, i.e. the space of smooth functions with compact support in~$E$.
		
	\end{notation}
	
	\vskip 1cm

	\begin{notation}[\textbf{Black-Scholes Bilinear Form}] $\, $\\

 	For any pair~\mbox{$(u,v)\, \in\, \mathcal{D}(\R_+)\times\mathcal{D}(\R_+)$}, set:

		\begin{align*}
		B(u,v)&=\displaystyle\int_{\R_+} \frac{\sigma^2 x^2}{2} \frac{\partial u}{\partial x} \frac{\partial v}{\partial x} dx + \int_{\R_+} \left(\sigma^2-r\right) x \frac{\partial u}{\partial x}\, v \,dx +\int_{\R_+} r\,u\,v\, dx 
		\, \cdot
		\end{align*}
		
\noindent and note $B(u,u)=B(u)$.

	\end{notation}
	
	\vskip 1cm

	\begin{pte}{\ }\\
		
  The bilinear form $B(\cdot ,\cdot )$ is non-symmetric.
		
	\end{pte}
	
	\vskip 1cm
	
 Define its symmetric part, for any pair~\mbox{$(u,v)\, \in\, \mathcal{D}(\R_+)\times\mathcal{D}(\R_+)$}, through:
	
	\begin{align*}
	\tilde{B}(u,v)&=\displaystyle \frac{1}{2}\, \left (B(u,v)+B(v,u)\right )\\
	&=\displaystyle \int_{\R_+} \frac{\sigma^2 x^2}{2} \frac{\partial u}{\partial x} \, \frac{\partial v}{\partial x} dx + \frac{1}{2}\, \int_{\R_+} \left(\sigma^2-r\right) \,x \, \left( \frac{\partial u}{\partial x}\, v  + u \,\frac{\partial v}{\partial x}\right) \,dx +\int_{\R_+} r\,u\,v\, dx \\
	&=\displaystyle \frac{\sigma^2}{2}\int_{\R_+} x^2\,  \frac{\partial u}{\partial x} \frac{\partial v}{\partial x} \, dx +\frac{3r-\sigma^2}{2}\int_{\R_+} u\,v\, dx 
	\end{align*}

	\vskip 1cm		
	
	\begin{notations}{\ }\\
		
		\noindent Set
		
		$$ V= \left \lbrace  v \, \in \, L^2(\R_+) \quad ,  \quad x\, \displaystyle \frac{\partial v}{\partial x}\, \in \, L^2(\R_+) \right\rbrace \quad , \quad 
		W= \left \lbrace  v \, \in \, L^2(\R_+) \quad ,  \quad x^2\, \displaystyle \frac{\partial^2 v}{\partial x^2}\, \in \, L^2(\R_+) \right\rbrace  \, \cdot $$
		
	\end{notations}

	\vskip 1cm

\newpage
	
	\begin{pte}{\ }\\

The space
		
		$$ V= \left \lbrace  v \, \in \, L^2(\R_+) \quad ,  \quad x\, \displaystyle \frac{\partial v}{\partial x}\, \in \, L^2(\R_+) \right\rbrace $$
		
		\noindent endowed with the inner product
		
		$$ (u,v) \mapsto \left\langle u,v \right\rangle_V = \left\langle u,v \right\rangle_{L^2(\R_+)} + \left\langle x\frac{du}{dx},x\, \displaystyle \frac{dv}{dx}\right\rangle_{L^2(\R_+)} \, \cdot $$
		
		\noindent is a Hilbert space.
		
	\end{pte}
	
	\vskip 1cm
	
	
	\begin{definition}[\textbf{Black-Scholes Weak Formula}] $\, $\\
		
		\noindent The Black-Scholes variational formula reads: find $u\in C([0,T];L^2(R_+))\cap L^2((0,T);V)$ such that $\partial_t u\in L^2((0,T);V^{\star})$, satisfying:
		
\begin{align*}
\begin{cases}
B(u,v)&=\displaystyle \frac{d}{dt} \, 	\displaystyle \int_{\R_+} u(t,x) \, v(x) \, dx \, \quad , \quad \forall v\, \in\, \mathcal{D}(\R_+)\\
u(T,x)&=h(x)
\end{cases}
\end{align*}

\noindent where $V^{\star}$ is the dual space of $V$.
	\end{definition}
	
	\vskip 1cm
	
	\begin{lemma}[\textbf{Black-Scholes Operator}] $\, $\\
		
 There exists a unique bounded linear operator~\mbox{$\mathcal{BS} \, : \, V\rightarrow V'$}, which will be called Black-Scholes operator, such that:
		
		$$ \forall\, (u,v)\,\in\, V^2:\quad B(u,v)=\left\langle \mathcal{BS}(u),v\right\rangle_{L^2(\R_+)}  \, \cdot$$
		
	\end{lemma}
	
	\vskip 1cm	
	
	\begin{definition}{\ }\\

		Recall that the first derivative is always bounded in the Hilbert-Sobolev space~\mbox{$H^1(\R_+)$}, and that the second derivative is always bounded in the Hilbert-Sobolev space~\mbox{$H^2(\R_+)$}.\\

 The Black-Scholes operator is given, for any~$v$ in~$W$, by
		
		$$ \mathcal{BS}(v)= -\displaystyle\frac{x^2\sigma^2}{2} \frac{\partial^2 v}{\partial x^2} - r \,x \, \frac{\partial v}{\partial x} + r\, v  \, \cdot$$
		
	\end{definition}
	
	\vskip 1cm
	
	\begin{lemma}{\ }\\ 
		
 The set
		
		$$ W= \left\{ v\,  \in V \quad , \quad x^2\, \displaystyle \frac{\partial^2 v}{\partial x^2}\, \in \, L^2(\R_+) \right\}$$
		
		\noindent is dense in~$V$.
		
	\end{lemma}
	
	\vskip 1cm
	
	\begin{lemma}[\textbf{Black-Scholes Weak Solution}] $\, $\\
		
 For $h$ in~$L^2(\R_+)$, the Black-Scholes problem has a unique weak solution.
	\end{lemma}
	
	\vskip 1cm
	
\subsection{An Interpretation of the Generalized Measure Black-Scholes Model}	
	
	\hskip 0.5cm The generalized measure Black-Scholes equation, for a probability measure $\mu$, can be written, by choosing $v=u$, as

	\begin{align*}
	 \frac{\partial }{\partial t}\mathbb{E}\left( |u|^2 \right) & =\frac{2}{\mathbb{M}}\, B\left(u\right) 
	\end{align*}
	
	\noindent where $\mathbb{M}=M-L$, while~$\mathbb{E}\left( . \right)$ is the expected value and $B(\cdot,\cdot)$ denotes the involved Black-Scholes bilinear form. This equation can be understood in the following way: the expected value depreciation of $u^2$ (and, hence, of~$u$, since it is positive) is proportional to the Black-Scholes energy $B(u)$. This implies a law given by the classical Black-Scholes theory which holds on \emph{average} and another induced by investors' perception of reality and reflected by $\mu$.\\
	
	A direct implication of this vision is a local dependence of the option price dynamics on the probability measure $\mu$, which can be interpreted as a$\backslash$n \textbf{confidence$\backslash$uncertainty} measure.

	\vskip 1cm
	
	\section{Non-Symmetric Dirichlet Forms and Generalized Measure Black-Scholes Operators \label{Non-Symmetric Dirichlet forms and Generalized Measure Black-Scholes Operator}}
	
	
\hskip 0.5cm Given two strictly positive numbers~$L<M$, if one replaces~$\R_+$ with $\overline{\mathcal{M}}=\left[L,M\right]$ (where $\mathcal{M}=\left]L,M\right[$) in the generalized measure Black-Scholes model, it does not result in a dramatic effect on the mathematical or economical foundations of the model from the following two points of view:

\begin{enumerate}
\item \textbf{Financial stability:} One can suppose, in \textbf{short run}, boundedness of the underlying financial instrument price.

\item \textbf{Numerical analysis:} It is well known that infinite boundaries are replaced with finite ones for numerical simulations (See \cite{Kangro2001} for error estimates).

\end{enumerate} 

In section \ref{Underlying asset price bounds}, a tolerance level associated with the choice of the bounds $L$ and $M$ is calculated.

\vskip 1cm
	
	\begin{notations}{\ }\\
		
Let's introduce the following two spaces:
		
\begin{align*}
V_{\mathcal{M}}&= \left \lbrace  v \, \in \, L^2(\mathcal{M}) \quad ,  \quad x\, \displaystyle \frac{\partial v}{\partial x}\, \in \, L^2(\mathcal{M}) \right\rbrace \quad ,\\ 
L^2_{\mu}(\mathcal{M})&= \left \lbrace  v  \quad ,  \quad \int_L^M v^2 \,d\mu<\infty \right\rbrace \quad \cdot
\end{align*}
		
  The dual space of~\mbox{$ V_{\mathcal{M}}$} will be denoted by~\mbox{$ V_{\mathcal{M}}^\star$} and the closure of $D(\mathcal{M})$ in $V_{\mathcal{M}}$ by $V_{0,\mathcal{M}}$.
		
	\end{notations}

\vskip 1cm

\begin{proposition}{\ }\\
		
  The space~$\mathcal{D}(\mathcal{M})$ is dense in~$L^2_{\mu}(\mathcal{M})$.
		
	\end{proposition}
	
	\vskip 1cm
		
 As in~\cite{Hu2006}, the fundamental conditions under which the solution exist are obtained thanks to the following assumptions:
	
	\vskip 1cm
	
	\begin{assumptions}{\ }\\
		
		\label{ExistenceCondition}
		\noindent For any~$u$ in~$\mathcal{D}(\mathcal{M})$, there exists a positive constant $C_0$:
		
\begin{align}
\|  u \| _{L^2_{\mu}(\mathcal{M})} &\varleq C_0  \left \|  x \, \displaystyle \frac{\partial u}{\partial x} \right \| _{L^2(\mathcal{M})} \, , &\text{(Continuous injection condition)}\\
\sigma^2 &< 4r \, \cdot  &\text{(Coercivity condition)}
\end{align}

	\end{assumptions}
	
	\vskip 1cm
	
	\begin{proposition}{\ }\\
		
  Under the first condition in assumption~\ref{ExistenceCondition}, there exists a unique~\mbox{$L^2_{\mu}(\mathcal{M})$}-representative $\tilde{u}$ of each equivalence class of functions~$u$ in~\mbox{$V_{\mathcal{M}}$} such that the above condition holds. There also exists a~\mbox{$\mathcal{D}(\mathcal{M})$} sequence~\mbox{$(u_n)_{n\in \N}$} which converges towards~$\tilde{u}$ both in~\mbox{$V_{\mathcal{M}}$} and in~\mbox{$L^2_{\mu}(\mathcal{M})$}, since~\mbox{$\mathcal{D}(\mathcal{M})$} is dense in~\mbox{$V_{\mathcal{M}}$} and~\mbox{$L^2_{\mu}(\mathcal{M})$} \cite{Kilpelainen1994}.
	\end{proposition}

	\vskip 1cm
	
  Consider the bilinear form~\mbox{$B(\cdot,\cdot)$} with domain~$\text{dom}\, (B)$ on the Hilbert space~\mbox{$ V_{\mathcal{M}}$}. We introduce  the bilinear form:
	
	$$B^{\star}(\cdot, \cdot)=B(\cdot, \cdot)+\left\langle\cdot, \cdot\right\rangle_{L^2_{\mu}(\mathcal{M})}$$
	
	\noindent and the symmetric one:
	
	$$\tilde{B}^{\star}(\cdot, \cdot)=\tilde{B}(\cdot, \cdot)+\left\langle\cdot, \cdot\right\rangle_{L^2_{\mu}(\mathcal{M})} \, \cdot$$
	
  We refer to~\cite{Ma1992} for further details on the theory of non-symmetric Dirichlet forms.
	
	\vskip 1cm
	
	\begin{definition}[\textbf{Symmetric Closed Form}] $\, $\\
		
  A pair $(B, \text{dom}\, (B))$ is a \textbf{symmetric closed form} (on $H$) if $\text{dom}\, (B)$ is a dense linear subspace of $H$ and $B \, : \,  \text{dom}\, (B) \times \text{dom}\, (B) \, \rightarrow \R$ is a positive definite bilinear which is symmetric and closed on $H$ (i.e., $\text{dom}\, (B)$ is complete with respect to the norm~\mbox{${B^{\star}(\cdot, \cdot)}^{\frac{1}{2}}$}). 
	\end{definition}
	
	\vskip 1cm
	
	\begin{definition}[\textbf{Sector Condition}] $\, $\\
		
  Let us denote by~$B$ a bilinear form on the Hilbert space $H$, and by~$\text{dom}\, (B)$ its domain. The pair~\mbox{$(B, \text{dom}\, (B))$} is said to satisfy:
		
		\begin{itemize}
			\item[\emph{i}.] \textbf{The weak sector condition} if there exists~\mbox{$K > 0$} such that: 
			
\begin{align}
\label{eqn:weaksectorcondition}
\forall \, (u, v)\, \in \, \text{dom}\, (B)\times \text{dom}\, (B): \quad  |B^{\star}(u,v)| \varleq K \, \sqrt{ B^{\star}(u,u) \, B^{\star}(v,v)} \cdot
\end{align}
			
			\item[\emph{ii}.] \textbf{The strong sector condition} if there exists~\mbox{$  K > 0$} such that

			$$\forall \, (u, v)\, \in \, \text{dom}\, (B)\times \text{dom}\, (B): |B(u,v)| \varleq K \, \sqrt{ B(u,u) \, B(v,v) } \, \cdot$$
		\end{itemize}
		
	\end{definition}
	
	\vskip 1cm
	
	\begin{remark}{\ }\\
		
	  A coercive continuous bilinear form satisfies both conditions.
	\end{remark}
	
	\vskip 1cm
	
	\begin{definition}[\textbf{Coercive Closed Form}]$\, $\\
		
  A pair~\mbox{$(B, \text{dom}\, (B))$} will be called a \textbf{coercive closed form} (on $H$) if $\text{dom}\, (B)$ is a dense linear subspace of $H$ and ~\mbox{$B \, : \,  \text{dom}\, (B) \times \text{dom}\, (B) \, \rightarrow \R$} is a bilinear form such that the following two conditions hold:
		
		\begin{itemize}
			
			\item[\emph{i}.] Its symmetric part~\mbox{$(\tilde{B},\text{dom}\, (B))$} is a symmetric closed form on~$H$.
			
			\item[\emph{ii}.] ~\mbox{$(B,\text{dom}\, (B))$} satisfies the weak sector condition inequality~\mbox{$i$.} given in  (\ref{eqn:weaksectorcondition}). 
		\end{itemize}
		
	\end{definition}
	
	\vskip 1cm
	
	\begin{definition}[\textbf{Symmetric Vs Non-Symmetric Dirichlet Form}] $\, $\\

 A coercive closed form~\mbox{$(B,\text{dom}\, (B))$} on~\mbox{$L^2_{\mu}(\mathcal{M})$}, for a given measure~$\mu$, will be called a \textbf{Dirichlet form} if, for any~$u $ in~$ \text{dom}\, (B)$, one has: 
		
		$$\tilde{u}\in \text{dom}\, (B) \quad \text{and} \quad 
		\left \lbrace 
		\begin{array}{ccc}
		B(u+\tilde{u},u-\tilde{u})&\geq &0\\
		B(u-\tilde{u},u+\tilde{u})&\geq &0\\
		\end{array}
		\right.$$

		\noindent where~\mbox{$\tilde{u}=\max(0,\min(u,1))$}. If~\mbox{$(B, \text{dom}\, (B))$} is in addition symmetric, this is equivalent to:
		
		\begin{align*}
		B(\tilde{u}, \tilde{u}) \varleq B(u, u) \, \cdot 
		\end{align*}
		
 $B$ will be called a \textbf{symmetric Dirichlet form}. 
		
	\end{definition}
	
	\vskip 1cm

	\begin{theorem}{\ \cite{Ma1992}}\\
		
  Let us denote by~\mbox{$(B, \text{dom}\, (B))$} a coercive closed form on $H$, and $J$ a continuous linear functional on $\text{dom}\, (B)$. Then, there exists a unique~\mbox{$u\, \in\,  \text{dom}\, (B)$} such that
		
		$$ \forall \, v \,\in\,\text{dom}\, (B):\quad B^{\star}(u,v)=J(v) \,\cdot$$
		
	\end{theorem}
	
	\vskip 1cm
	
	\begin{definition}[\textbf{Non-Symmetric Dirichlet Forms}] $\, $ \\
		
	A coercive closed form on~$H$,~$(B, \text{dom}\, (B))$, is said to be a \textbf{non-symmetric Dirichlet form} (on $H$), if there exists a one-to-one correspondence with a pair of bounded linear operators~\mbox{$(L,\tilde{L})$}:
		
		$$ \forall \, (u,v)\, \in \, \text{dom}\, (L)\times \text{dom}\, (L) : \quad B(u,v)=(-Lu,v)=(u,-\tilde{L}v)$$
		
		\noindent where~\mbox{$\text{dom}\, (L)$ } is the domain of~$L$. Also, $\text{dom}\, (L)$ is a dense subset of $\text{dom}\, (B)$.\\
		\noindent The operator $L$ (respectively $\tilde{L}$) is the generator of a strongly continuous contraction semi-group~\mbox{$(T_t)_{t>0}$} (respectively $(\tilde{T})_{t>0}$).
	\end{definition}

	\vskip 1cm
	
The following result follows from \cite{Achdou2005}.	
	
	\vskip 1cm
	
	\begin{proposition}[\textbf{Poincar\'e Inequality} ] $\, $ \\
		
  The space~$\mathcal{D}(\mathcal{M})$ is dense in~$V_{\mathcal{M}}$, and, for any~\mbox{$v\, \in\, \mathcal{D}(\mathcal{M})$},  the following inequality is satisfied:

		$$ \|  v \| _{L^2(\mathcal{M})} \varleq 2 \, \left \| x\, \displaystyle \frac{dv}{dx} \right\|_{L^2(\mathcal{M})} $$
		
  This inequality induces a second norm on $V_{\mathcal{M}}$, given, for any~$v$ in~$V_{\mathcal{M}}$, by:
		
		$$ | v |_{V_{\mathcal{M}}} = \left \|  x\frac{dv}{dx} \right \| _{L^2(\mathcal{M})} \, \cdot$$
		
	\end{proposition}
	
	\vskip 1cm
	
	\begin{proposition}{\textbf{(Continuity and G\r{a}rding Inequality)}\label{G\r{a}rding Inequality} }\\
		
  The bilinear form~$B(\cdot ,\cdot)$ is continuous on $V_{\mathcal{M}}$, and satisfies the G\r{a}rding inequality:

		$$
		\forall\, u\, \in\, V_{\mathcal{M}}: \quad 
		B(u)  \geq  \displaystyle \frac{\sigma^2}{2} | u |_{V_{\mathcal{M}}}^2 - \lambda \, \parallel u \parallel_{L^2(\mathcal{M})}^2   
		$$
		
		\noindent where~$\lambda=\displaystyle \frac{\left(\sigma^2-3r\right)}{2}$. Moreover,~$B(\cdot ,\cdot)$ is coercive under the second assumption \ref{ExistenceCondition}.
		
	\end{proposition}
	
	\vskip 1cm

		\begin{proof}{\ }\\
 	For any pair~\mbox{$(u,v)\, \in\, \mathcal{D}(\mathcal{M})\times \mathcal{D}(\mathcal{M})$}, by using the Poincaré inequality, we obtain that

		\begin{align*}
		|B(u,v)|&=\left|\displaystyle\int_L^M \frac{\sigma^2 x^2}{2} \frac{\partial u}{\partial x} \frac{\partial v}{\partial x} dx + \int_L^M \left(\sigma^2-r\right) x \frac{\partial u}{\partial x}\, v \,dx +\int_L^M r\,u\,v\, dx \right|\\
&\varleq\frac{\sigma^2}{2} |u|_{V_{\mathcal{M}}} |v|_{V_{\mathcal{M}}} + \left(\sigma^2-r\right) |u|_{V_{\mathcal{M}}} \parallel v \parallel_{L^2(\mathcal{M})} + \,r\,\parallel u \parallel_{L^2(\mathcal{M})}\parallel v \parallel_{L^2(\mathcal{M})} \\
&\varleq C_1 \, |u|_{V_{\mathcal{M}}} |v|_{V_{\mathcal{M}}}
		\end{align*}

where $C_1=\displaystyle 2r + \frac{5\sigma^2}{2}$. For the coercivity, we use again Poincaré inequality:

\begin{align*}
 B(u) &=\displaystyle\frac{\sigma^2}{2}| u |_{V_{\mathcal{M}}}^2  + \int_L^M \left(\sigma^2-r\right) x \frac{\partial u}{\partial x}\, u \,dx + r\parallel u\parallel_{L^2(\mathcal{M})}^2\\
 &=\frac{\sigma^2}{2}| u |_{V_{\mathcal{M}}}^2 - \frac{\left(\sigma^2-3r\right)}{2} \parallel u\parallel_{L^2(\mathcal{M})}^2\\
&\geq C_2 \, | u |_{V_{\mathcal{M}}}^2\\
\end{align*}

where $C_2=\displaystyle 6r-\frac{3\sigma^2}{2}$.
		
	\end{proof}

	\vskip 1cm
	
	\begin{definition}{\ }\\
		
  We introduce the mapping $\iota \, : \,  V_{\mathcal{M}} \rightarrow L^2_{\mu}(\mathcal{M})$ by
		
		$$\iota (u)= \bar{u}$$
		
		\noindent where $\bar{u}$ is the unique $L^2_{\mu}(\mathcal{M})$-representative of $u$, along with the closed set:
		
		$$ \mathcal{N}= \left\{ v \in   V_{\mathcal{M}} \, : \, \|    v \| _{L^2_{\mu}(\mathcal{M})} =0 \right\} \, \cdot $$
		
	\end{definition}
	
	\vskip 1cm
	
	\begin{theorem}{\textbf{The Black-Scholes Non-Symmetric Dirichlet Form} \\}
		
	  Under the assumption~\ref{ExistenceCondition}:

		\begin{enumerate}
			\item[ {i}.] $\text{dom}(B)= V_{\mathcal M}$ is dense in~\mbox{$L^2_{\mu}(\mathcal{M})$}.
			
			\item[ {ii}.]$\left (\tilde{B}^{\star},{V_{\mathcal M}}\right )$ is a Hilbert space.
			
			\item[ {iiii}.]$(B,\text{dom}(B))$ is a (non-symmetric) Dirichlet form.
		\end{enumerate}
		
	\end{theorem}

	\vskip 1cm

	\begin{proof}{\ }\\
		\begin{itemize}
			\item Let us consider a sequence~\mbox{$(u_n)_{n\in\N}$} of~\mbox{$\mathcal{D}(\mathcal{M})$}, which converges towards~\mbox{$u$} in~\mbox{$L^2_{\mu}(\mathcal{M})$}. \\
			
			We then consider two sequences,~\mbox{$(a_n)_{n\in\N}\, \in\, {\mathcal{N}}^\N$}, and~\mbox{$(b_n)_{n\in\N}\, \in\, {V_{\mathcal M}}^\N$} such that, for any natural integer~$n$,

			$$u_n = a_n + b_n \, \cdot$$
			
			Then, the sequence~\mbox{$(b_n)_{n\in\N}$} converges to~$u$ in~\mbox{$L^2_{\mu}(\mathcal{M})$}.\\
			
			\item Under the continuous injection condition (1) in assumption~\ref{ExistenceCondition}, the induced norm ${\tilde{B}^{\star}(\cdot, \cdot)}^{\frac{1}{2}}$ is equivalent to the norm~\mbox{$|\cdot |_{V_{\mathcal M}}$}. Hence,  $(B^{\star},\text{dom}(B))$ is complete.
			\item It follows from the coercivity of $B$ that:
			
			\begin{align*}
			0 \varleq C_2  \, |  u^2-\tilde{u}^2| ^2_{V_{\mathcal M}} \varleq B(u\pm\tilde{u},u\mp\tilde{u}) \, \cdot
			\end{align*}
		\end{itemize}
	\end{proof}
	
	\vskip 1cm
	
	\begin{theorem}[\textbf{Generalized Measure Black-Scholes Operator}] $\, $ \\
		
  Under the assumption~\ref{ExistenceCondition}, there exists a bounded linear operators $\mathcal{BS}_{\mu}$, that we will call~\textbf{generalized measure Black-Scholes operator}, such that, for any pair~ \mbox{$(u,v) \,\in\, \text{dom}(\mathcal{BS}_{\mu}) \times  \text{dom}(B)$}:
		
		$$ B(u,v)=\left\langle \mathcal{BS}_{\mu}(u), v\right\rangle_{L^2_{\mu}(\mathcal{M})}$$
		
 Moreover, we will say that $u \in \text{dom}(\mathcal{BS}_{\mu})$ and $\mathcal{BS}_{\mu}(u)=f$ if and only if
		
		$$ B(u,v)=\displaystyle \int_{\mathcal{M}} f\, v \, d\mu \, \quad , \quad \forall v\in V_{0,\mathcal M} $$
		
	\end{theorem}
	
	\vskip 1cm
	
\begin{remark}{\ \label{Black-Scholes operator continuity} }\\

The generalized measure Black-Scholes operator is bounded from $V_{\mathcal{M}}$ to $V_{\mathcal{M}}^{\star}$ since, $\forall v \in V_{0,\mathcal{M}}$

\begin{align*}
\left| \left\langle \mathcal{BS}_{\mu}(u), v\right\rangle_{L^2_{\mu}(\mathcal{M})} \right|&=  \left| B(u,v) \right|\\
&\varleq C_1 \, |u|_{V_{\mathcal{M}}} |v|_{V_{\mathcal{M}}}
\end{align*}

\noindent by continuity of $B(\cdot,\cdot)$.

\end{remark}

	\vskip 1cm
	
	\begin{notations}[\textbf{Sobolev Spaces}]$\ $ \\
		
		Given a continuous subset~$E$ of~$\R$,~$k\, \in\, \N$, and~$p \geq 1$, we recall that the classical Sobolev spaces on~$E$ are
		
		$$W^{k}_p \left ( E  \right)=
		\left \lbrace f \,\in\, L^p  \left ( E \right) \, , \, 
		\forall\, j \varleq k\, : \, f^{(j)}  \,\in\, L^p  \left ( E \right) \right \rbrace   $$
		
		\noindent and

		$$H^k \left ( E  \right)=W^{k}_2 \left ( E  \right)=
		\left \lbrace f \,\in\, L^2  \left ( E \right) \, , \, 
		\forall\, j \varleq k\, : \, f^{(j)}  \,\in\, L^2  \left ( E \right) \right \rbrace  \, \cdot$$
		
 The subspace~\mbox{$H_0^k $} of functions which vanish on the boundary~$\partial E$ is
		$$H_0^k \left ( E  \right)=
		\left \lbrace f \,\in\, L^2  \left ( E \right) \, , \, f_{|\partial E}=0\,  \text{and }
		\forall\, j \varleq k\, : \, f^{(j)} \,\in\, L^2  \left ( E \right) \right \rbrace  \, \cdot$$

	\end{notations}
	
	\vskip 1cm
	
  It directly comes form the abstract theory of partial differential equations \cite{Zeidler1990}, \cite{Wolka1987}, \cite{Lions1968} that:
	
	\vskip 1cm
	
	\begin{theorem}[\textbf{Generalized Measure Black-Scholes Weak Solution}] $\, $\\
		
  Let us define the Gelfand triple (or equipped Hilbert space)~\mbox{$  V_{\mathcal{M}}\subset L^2_{\mu}(\mathcal{M})\subset   V_{\mathcal{M}}^\star$}. For~$h$ in~\mbox{$L^2_{\mu}(\mathcal{M}) $}, the generalized measure Black-Scholes problem admits, under the assumptions \ref{ExistenceCondition}, a unique weak solution. Moreover, for~\mbox{$k \geq 1$}, the solution map:
		
		\begin{align*}
		L^2_{\mu}(\mathcal{M}) & \longrightarrow W^{k}_2 \left ([0,T] ;  V_{\mathcal{M}} \right ) \\
h(x) & \longrightarrow u(t,x) \qquad , \qquad x \in \mathcal{M} \quad \text{and} \quad  t \in [0,T]\\
		\end{align*}
		
		\noindent is continuous.
	\end{theorem}
	
	\vskip 1cm
	
	\section{The Self-Similar Black-Scholes Operator - a Pointwise Formula\label{The Self-Similar Black-Scholes Operator - a Pointwise Formula}}

\vskip 1cm

	\begin{definition}[\textbf{Self-Similar Measure on a Real Interval, Associated to a Set of Contractions~\cite{StrichartzLivre2006}}] $\, $\\
	
		\label{SelfSimilarMeasureOnM}		
We hereafter consider the particular case where the real interval~\mbox{$\mathcal{M} =\left]L,M\right[\subset\R_+$} is self-similar with respect to the family of contractions $\{f_1,f_2\}$, and where~$f_1$ and~$f_2$ are defined, for any real number~$x$, by:
		$$
		f_1(x)=\displaystyle \frac{1}{2}\,(x+L)\quad  , \quad 
		f_2(x)=\displaystyle \frac{1}{2}\,\left(x+M\right) \, \cdot $$
  
A measure~$\mu$ with full support on~$\overline{\mathcal{M}} $ will be called~\textbf{self-similar measure} on~$\overline{\mathcal{M}}$, relative to the set of contractions~\mbox{$\left (f_1,f_2\right)$} if, given a family of strictly positive weights~\mbox{$\left( \mu_1\, , \mu_2\right)$} such that
		
		$$ \mu_1+\mu_2=1 \, ,$$
		
		\noindent one has, then,
		
		$$ \mu= \displaystyle \mu_1\,\mu\circ f_1^{-1}+  \mu_2\,\mu\circ f_2^{-1} \, \cdot $$
		
	\end{definition}
	
	\vskip 1cm

\hskip 0.5cm Let us consider~\mbox{$u \, \in \, \text{dom}(\mathcal{BS}_{\mu})$} where $\mu$ is a self-similar measure according to definition \ref{SelfSimilarMeasureOnM}. $\mathcal{BS}_{\mu}$ is now called \textbf{the self-similar Black-Scholes operator} and we set:~\mbox{$\mathcal{BS}_{\mu}(u)=f   $}.\\
	
 In order to compute the explicit formula of $\mathcal{BS}_{\mu}$, we set $\mathbb{M}=M-L$ and we recall the self similar construction of $\mathcal{M}$:
	
	$$ \overline{\mathcal{M}}= \displaystyle f_1 (\overline{\mathcal{M}}) + f_2 (\overline{\mathcal{M}}) \, \cdot$$

	\vskip 1cm

	\begin{definition}[\textbf{Prefractal Graph Approximation}] $\, $\\
		
  We denote by~$V_0$ the ordered set of the (boundary) points:
		
		$$\left \lbrace L,M\right \rbrace$$
		
 We build the graph~${{\cal M}_0}$ by connecting the two extremities of~$V_0$.\\
		
	  For any strictly positive integer~$m$, we set:
		
		$$V_m =f_1 \left (V_{m-1}\right ) + f_2 \left (V_{m-1}\right ) \, \cdot $$
		
	 The set of points~$V_m$, where consecutive points are connected, will be denoted by~$  {{\mathcal M}_m}$.\\
		
The set~$V_m$ is called the {set of vertices} of the graph~$  {\mathcal M}_m$. By extension, we will write that
		
		$${\mathcal M}_m=f_1 \left ({\mathcal M}_{m-1}\right ) + f_2 \left ({\mathcal M}_{m-1}\right ) \, \cdot$$
		
	\end{definition}

	\vskip 1cm
	
	One can prove that the sequence $\{V_m\}_{m\in\N}$ is increasing and its limit is dense in $\mathcal{M}$ (\cite{Kigami2001}).
	
	\vskip 1cm
	
	\begin{proposition}{\ }\\
	
		Given a natural integer~$m$, we will denote by~$\mathcal{N}_m$ the number of vertices of the graph ${\mathcal{M}}_m$. One has:
		
		$$\mathcal{N}_0  =2$$
		
		\noindent and, for any strictly positive integer~$m$:
		
		$$\mathcal{N}_m=2^m+1 \, \cdot$$
	\end{proposition}
	
	\vskip 1cm

We recall the following definitions:	
	
	\vskip 1cm
		
	\begin{definition}[\textbf{Word}] $\, $\\

  Given a strictly positive integer~$m  $, we will call \textbf{number-letter} any integer~${\mathcal{W}}_i$ of~\mbox{$\left \lbrace 1, 2  \right \rbrace $}, and \textbf{word of length~$|{\mathcal{W}}|=m$}, on the graph~$ { \mathcal{M}}_m $, any set of number-letters of the form:

		$${\mathcal{W}}=\left ( {\mathcal{W}}_1, \hdots, {\mathcal{W}}_m\right) \, \cdot$$
		
  We will write:
		
		$$f_{\mathcal{W}}= f_{{\mathcal{W}}_1} \circ \hdots \circ  f_{{\mathcal{W}}_m} \, \cdot$$
		
	\end{definition}
	
	\vskip 1cm
	
	\begin{definition}[\textbf{Addresses}]$\, $\\
		
  Given a natural integer~$m$, and a vertex~$X$ of~${\cal M}_m$, we will call {address of the vertex~$X$} an expression of the form

		$$X= f_{ \cal W} \left (L\right ) \quad \text{or} \quad X =f_{ \cal W^{'}} \left (M\right )$$

		\noindent	where~$\cal W$  and~${\cal W}'$ denote words of length~$m$. The vertex~$X$ has thus a double address.

	\end{definition}
	
	\vskip 1cm

\begin{pte}[\textbf{Space of Harmonic Splines}] $\, $\\
	
  Given a strictly positive integer~$m$, we introduce the space of harmonic splines of order $m$, denoted by~\mbox{$\mathcal{H}^m(\left[L,M\right])$}, as the space of functions~\mbox{$\psi^m_X$},~\mbox{$X\,\in\, [L,M]$}, such that:
	
	$$\forall \, Y\, \in\, \mathcal{M}_m \quad \psi^m_X(Y)=\delta_{XY} \, \cdot $$
	
  For~\mbox{$k \, \in \left \lbrace  1,\hdots,2^m-1\right \rbrace $}, and~\mbox{$Y\,\in\, [L,M]$}:
	
	\begin{align*}
	\psi^m_{L+\frac{k\,\mathbb{M}}{2^m}}(Y)&=\begin{cases}  	\displaystyle \frac{2^m}{\mathbb{M}} (Y-L) - (k-1) \qquad L+\frac{(k-1)\mathbb{M}}{2^m} \varleq Y \varleq L+\frac{k\,\mathbb{M}}{2^m}\\ 
	-	\displaystyle\frac{2^m}{\mathbb{M}} (Y-L)+ (k+1) \qquad 	\displaystyle L+\frac{k\,\mathbb{M}}{2^m} \varleq Y \varleq L+\frac{(k+1)\,\mathbb{M}}{2^m}\\  0 \qquad \text{otherwise} \end{cases}
	\end{align*}
	
	\noindent and
	
	\begin{align*}
	\displaystyle 	\psi^m_L(Y) &=\begin{cases} -	\displaystyle \frac{2^m}{\mathbb{M}} (Y-L)+ 1 \qquad L \varleq Y \varleq L+\frac{\mathbb{M}}{2^{m}}\\ 0 \qquad \text{otherwise} \end{cases} \quad , \\ 
	\psi^m_M(Y)&=\begin{cases} 	\displaystyle \frac{2^m}{\mathbb{M}} (Y-M)+ 1 \qquad M-\frac{\mathbb{M}}{2^m} \varleq Y \varleq M\\ 0 \qquad \text{otherwise} \end{cases}
	\end{align*}

\end{pte}

\vskip 1cm

	\begin{proposition}[\textbf{Integration of Harmonic Splines}] $\, $\\
		
	  Let us consider a strictly positive integer~$m$. For~\mbox{$k \, \in \left \lbrace  1,\hdots,2^m-1\right \rbrace $}, we denote by~\mbox{${\mathcal{ V_{\mathcal{M}}}}_k$} and~\mbox{${\mathcal{W}}_k$} the unique indices such that
		
		$$ f_{{\mathcal{ V_{\mathcal{M}}}}_k}\left (\left[L,M\right]\right )=	\displaystyle \left[L+\frac{(k-1)\mathbb{M}}{2^m},L+\frac{k\,\mathbb{M}}{2^m}\right] \quad \text{and} \quad  
		f_{{\mathcal{W}}_k}\left (\left[L,M\right] \right )=	\displaystyle \left[L+\frac{k\,\mathbb{M}}{2^m},L+\frac{(k+1)\,\mathbb{M}}{2^m}\right] \, \cdot $$
		
  Then,
		
		$$
		\displaystyle  
		\int_L^M \psi^m_{L+\frac{k\,\mathbb{M}}{2^m}} \, d\mu  = {\mu_1}^{s_{{\mathcal{ V_{\mathcal{M}}}}_k}}{\mu_2}^{m+1-s_{{\mathcal{ V_{\mathcal{M}}}}_k}}+\mu_1^{s_{{\mathcal{W}}_k}+1}\, 
		\mu_2^{m-s_{{\mathcal{W}}_k}} $$
		\noindent and
		
		$$	\displaystyle 	\int_L^M \psi^m_{L} \, d\mu = {\mu_1}^{m+1} \quad \quad 
		\int_L^M \psi^m_{M} \, d\mu = {\mu_2}^{m+1} \, \cdot $$
		
	  In addition, if ${\mu_1}< 	\displaystyle \frac{1}{2}$:
		
		\begin{align*}
		\displaystyle 	\int_L^M \psi^m_{L} d\mu < \int_L^M \psi^m_{L+\frac{k\,\mathbb{M}}{2^m}} \, d\mu <
		\int_L^M \psi^m_{M} \, d\mu \, \cdot \\
		\end{align*}
		
	\end{proposition}
	\vskip 1cm

	\begin{pte}

	 Given a strictly positive integer~$m$, we set, for any integer~\mbox{$ k \, \in\, \left \lbrace 0, \hdots, 2^m \right \rbrace$}: 
		
		$$x_k=\displaystyle L+\left(k\,\frac{\mathbb{M}}{2^m}\right)  $$
		
		\noindent and, for any~$u \in \text{dom}(\mathcal{BS}_{\mu})$:
		
		\begin{align*}
		B(u,\psi^m_{x_k})&= \displaystyle\int_L^M \left(-\frac{\sigma^2 x^2}{2} \frac{\partial^2 u}{\partial x^2} - r x \frac{\partial u}{\partial x} + r\,u\right)\,\psi^m_{x_k}\, dx\\
		&=\lim_{m\rightarrow + \infty}\sum_{j=0}^{2^m} \,\mathcal{BS}_m\,(u(t,x_j))\,\psi^m_{x_k}\,\left(\frac{\mathbb{M}}{2^m}\right)\\
		&=\lim_{m\rightarrow + \infty} \,\mathcal{BS}_m\,(u(t,x_k))\,  \left(\frac{\mathbb{M}}{2^m}\right)\\
		\end{align*}
		
		\noindent where $\mathcal{BS}_m$ is the discrete Black-Scholes operator defined, for any~$t$ in~\mbox{$[0,T]$}, by:
		
		\begin{align*}
		\mathcal{BS}_m \, u(t,x_k)&= - \displaystyle\frac{\sigma^2}{2}\, x_k^2\left( \displaystyle \frac{u(t,x_{k+1})-2\,u(t,x_{k})+u(t,x_{k-1})}{\left(\frac{\mathbb{M}}{2^m}\right)^2}\right)\\
		& - r\, x_{k} \left( \frac{u(t,x_{k+1})-u(t,x_{k-1})}{2\left(\frac{\mathbb{M}}{2^m}\right)}\right)+r\,C(t,x_{k})\\
		&= -\frac{\sigma^2}{2}\, k^2\left( u(t,x_{k+1})-2\,u(t,x_{k})+u(t,x_{k-1})\right)\\
		& - r\, {k} \left( \frac{u(t,x_{k+1})-u(t,x_{k-1})}{2}\right)+r\,C(t,x_{k})\\
		\end{align*}

	  The mean value formula yields asymptotically
		
		\begin{align*}
		\displaystyle\int_L^M \mathcal{BS}_{\mu} u(x)\,\psi^m_{x_k} \, d\mu &\approx \mathcal{BS}_{\mu}u(x_k) \, \int_L^M \psi^m_{x_k} \, d\mu \, \cdot
		\end{align*}

	\end{pte}
	
	
	\vskip 1cm	

	\begin{theorem}[\textbf{Self-Similar Black-Scholes Operator Pointwise Formula}] $\ $\\
		
  Let us consider~\mbox{$u \, \in \, \text{dom}(\mathcal{BS}_{\mu})$}. Then, for any~\mbox{$x\, \in\, \overline{\mathcal{M}}$}, and any sequence~\mbox{$(x_m)_{m\in\N}$} of~\mbox{$V_m\setminus V_0$} which uniformly converges towards~$x$:
		
		\begin{align*}
		\mathcal{BS}_{\mu}(u)(x)= \displaystyle\lim_{m\rightarrow +\infty} 2^{-m} \, \left( \displaystyle \int_L^M \psi^m_{x_m} \, d\mu \right)^{-1} \,  \mathcal{BS}_m(u)(x_m) \, \cdot
		\end{align*}
	\end{theorem}
	
	\vskip 1cm
	
	\begin{proof}{\ }\\
	  The uniform convergence directly comes from the fact that
		
		$$
		\displaystyle\left(\frac{\mathbb{M}}{2^m}\right)\left(\displaystyle \int_L^M \psi^m_{x_m} d\mu \right)^{-1}  \mathcal{BS}_m(u)(x_m)  = C \, \frac{\displaystyle\int_L^M \mathcal{BS}_{\mu}\,\psi^m_{x_m} \, d\mu}{ \displaystyle \int_L^M \psi^m_{x_m} d\mu}\\
		=C\,\mathcal{BS}_{\mu}(u)(x) \, \cdot $$
		
  For~\mbox{$\mu_1=\mu_2= \displaystyle\frac{1}{2}$}, one recovers the classical Black-Scholes operator, which implies that: $C=\mathbb{M}$. 
	\end{proof}
	
	\vskip 1cm

	\section{Proof of the Assumptions}
	
	\hskip 0.5cm We hereafter consider the general case of non atomic finite measure $\mu$ with total mass $\mathbb{M}=M-L$ (In the case of self-similar measures, it suffice to multiply the measure by $\mathbb{M}$).
	
	\subsection{First assumption \ref{ExistenceCondition}}

	\begin{notation}[\textbf{Space of Weighted Continuous Functions}] $\, $\\
			
We will denote by~\mbox{$C(\overline{\mathcal{M}})$} the space of weighted continuous functions on~$\overline{\mathcal{M}}$ endowed with the norm

$$\|  u \| _{\eta,\infty}=\underset{x\, \in\,\overline{\mathcal{M}}}{\max}\,|x \, u(x)| \, \cdot $$
			
	\end{notation}
				
	\vskip 1cm
	
	\begin{proof}{\textbf{of the first assumption}\\}

	 Let us consider~$u \in V_{0,\mathcal{M}}$. On the one hand, we have that

\begin{align*}
		\left |x\, u(x) \right |&=	\left| \displaystyle \int_L^x\, (su(s))' \,ds \right |\\
		&=	\left|\int_L^x u(s) \,ds+\int_L^x su'(s) \,ds \right |\\
		&\varleq \sqrt{\mathbb{M}}\left( \|  u\| _{L^2}+ |u|_{  V_{\mathcal{M}}}\right)\\
		&\varleq \sqrt{\mathbb{M}} \|  u\| _{ V_{\mathcal{M}}}
		\end{align*}

		We deduce the continuity of the injection $\iota \, :\, (V_{\mathcal{M}},\| \cdot \| _{ V_{\mathcal{M}}}) \rightarrow (C_{\eta}(\overline{\mathcal{M}}),\|  \cdot \| _{\eta,\infty})$. On the other hand, for~$u \in C_{\eta}(\overline{\mathcal{M}})$, one also has that
		
		$$ \|  u \| _{L^2_{\mu}(\mathcal{M})}=\left(\int_L^M \frac{1}{x^2} \left( x\,u\right)^2 \,d\mu\right)^{\frac{1}{2}} \varleq \|  u \| _{\eta,\infty}\, \left(\int_L^M \frac{1}{x^2} \,d\mu\right)^{\frac{1}{2}} \varleq \|  u \| _{\eta,\infty}\, \frac{\mu(\overline{\mathcal{M}})^{\frac{1}{2}}}{L} $$
		
		\noindent  the injection~\mbox{$\iota \, : \, (C_{\eta}(\overline{\mathcal{M}}),\|  \cdot \| _{\eta,\infty}) \rightarrow (L^2_{\mu}(\mathcal{M}),\| \cdot \| _{L^2_{\mu}(\mathcal{M})})$} is continuous, so we obtain that
		
$$ \|  u \| _{L^2_{\mu}(\mathcal{M})} \varleq C_0 \,\|  u \| _{V_{\mathcal{M}}}\, $$

\noindent for $C_0=\displaystyle \frac{\sqrt{\mathbb{M}\,\mu(\overline{\mathcal{M}})}}{L}=\frac{\mathbb{M}}{L}$.

	\end{proof}

\vskip 1cm
	
	\subsection{Commentary on the second assumption \ref{ExistenceCondition}}
	
\hskip 0.5cm The assumption $4r > \sigma^2$ is not that restrictive as it may seem in the first sight, for example, if we give a look to sample from Vance L. Martin data~\cite{Martin2005}. The sample consist of~\mbox{$\mathbf{N}=269$} observations on the European call options written on the~\mbox{S$\&$P$500$} stock index on the~\mbox{$4^{\textrm{th}}$} of April, $1995$. One can calculate (see \cite{RianeInv})

\begin{itemize}
		\item  The interest rate $r=0.0591$.
		\item The volatility $\sigma=0.076675$.		
\end{itemize}

\noindent which means that

\begin{align*}
4r=0.2364 \gg 0.00587906=\sigma^2 \, \cdot\\
\end{align*}

The second assumption could also be replaced by the following condition: for any~$u$ in~$\mathcal{D}(\mathcal{M})$,

\begin{align*}
\|  u \| _{L^2(\mathcal{M})} \varleq C_3 \|  u \| _{L^2_{\mu}(\mathcal{M})} \, \cdot \\
\end{align*}
		
Given~$u$ in~\mbox{$\text{dom}\, (B)$}, set, in the spirit of~\cite{Wolka1987}:
		
		$$\tilde{\lambda}=\lambda\,C_3 \quad ,\quad  w(t,x)=e^{-\tilde{\lambda}\, t}u(T-t,x)$$
		
		\noindent  where $\lambda$ is G\r{a}rding's inequality constant in proposition \ref{G\r{a}rding Inequality}. Then $\forall \, v\, \in \, \text{dom}\, (B)  $, with $w(0,x) =h(x)$, one has:

		$$ \displaystyle \int_L^M \frac{d}{dt}w(t,x)  \, v(x) \, d\mu  = - B(w,v)-\tilde{\lambda} \, \int_L^M  w(t,x) \,  v(x) \, d\mu =-\widehat{B}(w,v)  \quad 
		 \, \cdot
		$$
		
Without affecting the solution space, one obtains the continuity and the coercivity of the form $\widehat{B}$:
		
		\begin{align*}
		|\widehat{B}(u,v)|&\varleq 
		C \,  |u|_{V_{\mathcal M}}\, |v|_{V_{\mathcal M}} +\tilde{\lambda}\,  \parallel u \parallel_{L^2_{\mu}(\mathcal{M})}\,  \parallel v \parallel_{L^2_{\mu}(\mathcal{M})}\\
		&\varleq(C+\tilde{\lambda}\,  C_0)\,  |u|_{V_{\mathcal M}}\, |v|_{V_{\mathcal M}} 
		\end{align*} 
		
		\noindent and:
		
		\begin{align*}
		\widehat{B}(u,u)&\geq 
		\displaystyle \frac{\underline{\sigma}^2}{4} \, | u |_{V_{\mathcal M}}^2 - \lambda  \parallel u \parallel_{L^2(\mathcal{M})}^2  + \tilde{\lambda} \, \parallel u \parallel_{L^2_{\mu}(\mathcal{M})}^2 \\
		&\geq  \frac{\underline{\sigma}^2}{4} |\,  u |_{V_{\mathcal M}}^2
		\end{align*} 
	
	\vskip 1cm

\subsection{Underlying asset price bounds\label{Underlying asset price bounds}}

\hskip 0.5cm An important question arises: how to choose $M$ and $L$ ? one answer is to use the law of $S_t$, the underlying asset price, then choose $M(\alpha)$ (respectively $L(\alpha)$) using the rule:

$$ 1-\mathbb{P}\left(L(\alpha)\varleq S_t \varleq M(\alpha) \right)\varleq \alpha \quad , \quad 0\varleq t \varleq T $$

\noindent for some tolerance level $\alpha$.\\

One can deduce a similar rule in the case of boundary conditions. For example, in the case of a call option:

$$ \mathbb{P}\left( S_T > K \, | \, S_0= L(\alpha) \right)\varleq \alpha $$

\noindent and

$$ \mathbb{P}\left( S_T \varleq K \, | \, S_0= M(\alpha) \right)\varleq \alpha $$

\vskip 1cm
	
	\section{Numerical Simulation of Self-Similar European Options}

\hskip 0.5cm Let us consider as in section \ref{The Self-Similar Black-Scholes Operator - a Pointwise Formula}, the self-similar case, for an European options call, defined by the system:
	
	\begin{align*}
	\displaystyle \frac{\partial C}{\partial t}(t,S)  & = \mathcal{BS}_{\mu}(C)(t,S) & \forall \,  t\, \in \, \left[0,T\right], \,\forall \, S \, \in\,  \mathcal{M} \\
	C(T,S)&=h(S) & \forall \, S\,  \in \, \mathcal{M}\\
	C(t,L)&=0 & \forall \, t\, \in\,  \left[0,T\right]\\
	C(t,M)&=g(t)& \forall \, t\in \left[0,T\right] \\
	\end{align*}
	
	\noindent for a self-similar measure $\mu$ on $\overline{\mathcal{M}}$, under the condition
	
	$$ 4\,r > \sigma^2 \, \cdot$$
	
	\noindent where~\mbox{$h(S)=(S-K)^+$} and~\mbox{$g(t)= M - K\, \exp(-r(T-t))$}, and where the constant $\sigma$ is the volatility, $r$ the risk-free interest rate,~~$T$ the maturity of the option and $C$ represents the call price.\\
	
  We will use the following change of variables:~\mbox{$\tau=T-t$}, which leads to the following equation (with the same notations):
	
	\begin{align*}
	- \displaystyle \frac{\partial C}{\partial t}(t,S)  & = \mathcal{BS}_{\mu}(C)(t,S) & \forall \,t \,\in\, \left[0,T\right], \,\forall \, S\,  \in \, \mathcal{M} \\
	C(0,S)&=h(S) & \forall \,S \,\in\, \mathcal{M}\\
	C(t,L)&=0 & \forall \,t\,\in \,\left[0,T\right]\\
	C(t,M)&=g(t)& \forall \,t\,\in\, \left[0,T\right] \\
	\end{align*}
	
	\noindent for $g(t)=M - K\,\exp(-r\,t)$.
	
	\vskip 1cm
	
\begin{remark}{\ }\\

The results of section \ref{Non-Symmetric Dirichlet forms and Generalized Measure Black-Scholes Operator} still hold, if we write $\tilde{C}=C-\tilde{g}$, where $\tilde{g}(t,x)=\displaystyle (x-L) \left(\frac{M-K\,\exp(-r\,t)}{M-L} \right)$, and replace the space $V_{\mathcal{M}}$ by $V_{0,\mathcal{M}}$, then we solve the non homogeneous problem

$$ \int_L^M  \, \frac{\partial \tilde{C}}{\partial t} \, v \, d\mu + B(\tilde{C} ,v) = \int_L^M \frac{d \tilde{g}}{d t} \, v \,d\mu $$
 
\noindent applying abstract theory of partial differential equations \cite{Wolka1987}.

\end{remark}

 	\vskip 1cm
	
	\subsection{The Finite Difference Method}

	\hskip 0.5cm In the spirit of our previous work~\cite{RianeDavidFDMS}, we fix a strictly positive integer~$N $, and set:
	
	$$ \displaystyle{h =\displaystyle\frac{T}{N}}$$
	
  We will write, for a function $f$,~\mbox{$n\, \in\, \left \lbrace 0,\hdots,N\right \rbrace $} and~\mbox{$k\,\in\, \left \lbrace 0,\hdots,2^m\right \rbrace$}:
	
	$$ f(h,k) =  f(n\,h,L+k\,	\displaystyle \frac{\mathbb{M}}{2^m}) \, \cdot $$

  We use the Euler implicit scheme, for any integer~$n$ belonging to~$\left \lbrace 0, \hdots, N-1 \right \rbrace$:
	
	$$   \forall \,k\,\in \, \{0,\hdots,2^m\} \,:   \quad
	\displaystyle\frac{\partial C}{\partial t}(n,k)\approx \displaystyle\frac{1}{h}\,\left( C(n+1,k)-C(n,k)\right)
	$$
	
  The self-similar Black-Scholes operator for~$k=\{0,\hdots,2^m\}$ is approximated through 
	
	\begin{align*}
	\mathcal{BS}_{\mu} \, C(n,k)&\approx \displaystyle  \left( \frac{2^{-m}}{{\mu_1}^{s_{\mathcal{ V_{\mathcal{M}}}}}{\mu_2}^{m-s_{\mathcal{ V_{\mathcal{M}}}}+1}+ {\mu_1}^{s_{\mathcal{W}}+1}{\mu_2}^{m-s_{\mathcal{W}}} }\right) 
	\mathcal{BS}_m \, C(n,k)\\
	&\approx \displaystyle  \delta_m \,
	\mathcal{BS}_m \, C(n,k)\\
	\end{align*}
	
	\noindent where $\mathcal{BS}_m$ is the discretized Black-Scholes operator given by
	
	\begin{align*}
	\mathcal{BS}_m \, C(n,k)&= -\frac{\sigma^2}{2}\left(\frac{k\,\mathbb{M}}{2^m}\right)^2\left( \frac{C(n,k+1)-2\,C(n,k)+C(n,k-1)}{\left(\frac{\mathbb{M}}{2^m}\right)^2}\right)\\
	& - r\, \left(\frac{k\,\mathbb{M}}{2^m}\right) \left( \frac{C(n,k+1)-C(n,k-1)}{2\left(\frac{\mathbb{M}}{2^m}\right)}\right)+r\,C(n,k) \, \cdot \\
	\end{align*}
	
  For~\mbox{$0 \varleq n \varleq N-1$}, and~\mbox{$1\varleq k \varleq 2^m-1$}, we define the following scheme:
	
	$$\left ({\mathcal S}_{\mathcal BS}\right) \quad \left \lbrace \begin{array}{cccc}
	\displaystyle -\frac{\mathcal{C}_{h,m}(n+1,k)-\mathcal{C}_{h,m}(n,k)}{h}&=&\displaystyle \delta_m \,\mathcal{BS}_m\,\mathcal{C}_{h,m}(n,k)&   \\
	\mathcal{C}_{h,m}(n,0)&=&0&     \\
	\mathcal{C}_{h,m}(n,2^m)&=&g(n) \\
	\mathcal{C}_{h,m}(0,k)&=&h(k) \, \cdot &
	\end{array} \right.  $$
	
  For~\mbox{$n\,\in \, \left \lbrace 0, \hdots, N-1 \right \rbrace$}, we set:
	
	$$
	\mathcal{C}(n)  =\left(
	\begin{matrix}
	\mathcal{C}_{h,m}(n,1)\\
	\vdots\\
	\mathcal{C}_{h,m}(n,2^m-1)\\
	\end{matrix}
	\right)\\
	=\left(
	\begin{matrix}
	\mathcal{C}_{h,m}(n,f_{\mathcal{W}^1}(M))\\
	\vdots\\
	\mathcal{C}_{h,m}(n,f_{\mathcal{W}^{2^m-1}}(M))\\
	\end{matrix}
	\right)
	\quad , \quad \left \lbrace \mathcal{W}^1,\hdots,\mathcal{W}^{2^m-1}\right \rbrace  
	\, \in\, \left \lbrace 1,2\right \rbrace^{m}   \, \cdot$$
	
 We have the following recurrence relation:
	
	$$\mathcal{C}(n+1) = A \,\mathcal{C}(n)+ B(n)$$
	
	\noindent where the~$ {(2^m-1)}\times  {(2^m-1)}$ matrix~$A$ is given by:
	
	\begin{align*}
	A&=I_{2^m-1} - h\, 2^{-m} \, \Psi_m^{-1} \, BS_m\\
	\end{align*}
	
	\noindent  and where~$I_{2^m-1}$ denotes the~$ {(2^m-1)}\times  {(2^m-1)}$ identity matrix, ${\Psi}_{m}$ and $BS_m$ the~$ {(2^m-1)}\times  {(2^m-1)}$ matrices:
	
	\begin{align*}
	{\Psi}_m &= \left(
	\begin{matrix}
	\ddots &   &   & 0 \\
	& {\mu_1}^{s_{\mathcal{ V_{\mathcal{M}}}}}{\mu_2}^{m+1-s_{\mathcal{ V_{\mathcal{M}}}}}+{\mu_1}^{s_{\mathcal{W}}+1}{\mu_2}^{m-s_{\mathcal{W}}} &    \\
	0 &  &   & \ddots \\
	\end{matrix}
	\right)
	\end{align*}
	
	\begin{align*}
	{BS}_m &= \left(
	\begin{matrix}
	-\sigma^2 - r & \frac{\sigma^2}{2} + \frac{r}{2} & 0 & \hdots  & 0 & 0 \\
	2\,\sigma^2 - r & -4\sigma^2 - r & 2\,\sigma^2 +  r & \hdots  & 0 & 0 \\
	\vdots  & \vdots  & \vdots  &\ddots &\vdots  & \vdots  \\
	0 & 0 & 0 & \hdots & -(2^m-2)^2\sigma^2 - r & (2^m-2)^2\frac{\sigma^2}{2} + (2^m-2)\frac{r}{2}\\
	0 & 0 & 0 & \hdots  &  (2^m-1)^2\frac{\sigma^2}{2} - (2^m-1)\frac{r}{2} & -(2^m-1)^2\sigma^2 - r \\
	\end{matrix}
	\right)\\
	\end{align*}
	
	\noindent and
	
	\begin{align*}
	B(n) &= h\,2^{-m} \, {\mu_2}^{-(m+1)} \left((2^m-1)^{2}\,\frac{\sigma^2}{2}  +(2^m-1)\, \frac{r}{2}\right) \left(
	\begin{matrix}
	0\\
	\vdots\\
	0\\
	g(n)\\
	\end{matrix}
	\right)  \, \cdot
	\end{align*}
	
	\vskip 1cm

	\subsection{Numerical Analysis}
	
	\vskip 0.5cm
	
	\paragraph{The Scheme Error and Consistency}{\ }\\
	
	\vskip 0.25cm
	
  Let us consider a function~ $ V_{\mathcal{M}}$ defined on~$\mathcal{M}$. For any~integer~$n$ in~\mbox{$\left \lbrace 0, \hdots, N-1 \right \rbrace$}, and any~$X$ in~${\mathcal{M}}$:
	
	$$    
	\displaystyle\frac{\partial v}{\partial t}(nh,X)=\displaystyle\frac{1}{h}\,\left( v((n+1)h,X)-v(nh,X)\right)+{\mathcal O}(h)
	$$
	
 As in~\cite{RianeDavidFDMS}, for any strictly positive integer~$m$, and any~$X$ in~
	\mbox{$V_m\setminus V_0$}, we can prove that:
	
	$$
	2^{-m}\, 		\displaystyle\left( \int_L^M \psi^m_{X} d\mu \right)^{-1}  \mathcal{BS}_m\,v(X)  = 
	\displaystyle \frac{\int_L^M \mathcal{BS}_{\mu}\,\psi^m_{X}\, d\mu}{		\displaystyle \int_L^M \psi^m_{X} \, d\mu}$$
	
	\noindent and that there exists a vertex $Z$ in the~$m$-cell $f_{\mathcal{W}}\left(\left[L,M\right]\right)$ containing $X$ such that
	
	\begin{align*}
	\left|\mathcal{BS}_{\mu} \,v(X) - 2^{-m}\, \left( \int_L^M \psi_X^{m}d\mu \right)^{-1} \mathcal{BS}_m \, v(X)\right| &= \left|\mathcal{BS}_{\mu} \, v(X) - \mathcal{BS}_{\mu}\, v(Z) \right|\\
	&\lesssim   \left| v(X) - v(Z) \right|\\
	&\lesssim   |X-Z|\\
	&\lesssim \left(\frac{1}{2}\right)^{ m}   \, \cdot 
	\end{align*}
	
\noindent using remark \ref{Black-Scholes operator continuity} and the uniform continuity of $v\in V_{\mathcal{M}}$. Thus,
	
	$$
	\mathcal{BS}_{\mu}\, v(X)  =  2^{-m}\,\left( \int_L^M \psi_X^{m}d\mu \right)^{-1} \mathcal{BS}_m \, v(X) + \mathcal{O}(2^{- m})   \, \cdot 
	$$
	
  The consistency error of our scheme is given by :
	
	\[
	\varepsilon^{h,m}_{n,k} = \mathcal{O}(h) + \mathcal{O}(2^{- m}) \quad 0\varleq n \varleq N,\, 0\varleq k \varleq 2^m  \, \cdot 
	\]
	
 We can check that the scheme is consistent:
	
	\[ \lim_{h\rightarrow 0, m\rightarrow\infty} \varepsilon^{h,m}_{n,k}=0  \, \cdot \] 
	
	\vskip 1cm
	
	\paragraph{Stability}{\ }\\
	
	\vskip 0.25cm
	
  We hereafter prove that the scheme is conditionally stable for the $\| \cdot  \| _{\infty}$ norm.
	
	\vskip 1cm
	
  Let us recall that, for~\mbox{$0\varleq n \varleq N$}, and~\mbox{$ 0\varleq k \varleq 2^m$}:
	
	\begin{align*}
	\displaystyle \mathcal{C}_{h,m}(n+1,k)&=  \mathcal{C}_{h,m}(n,k)\left(1 - h\,\delta_m \, \sigma^2\,k^2\right) \\
	&+\mathcal{C}_{h,m}(n,k+1) \left( h\,\delta_m \, \frac{\sigma^2}{2}\,k^2 +  h\,\delta_m \, \frac{r}{2}\, k \right)\\
	&+\mathcal{C}_{h,m}(n,k-1) \left( h\,\delta_m \, \frac{\sigma^2}{2}\,k^2 -  h\,\delta_m \, \frac{r}{2}\, k \right)\\
	&- \mathcal{C}_{h,m}(n,k) \,h\,\delta_m \, r \\
	&=\mathcal{C}_{h,m}(n,k)\left(1 - \alpha_k \right) +\mathcal{C}_{h,m}(n,k+1) \left( \frac{\alpha_k}{2} +  \beta_k \right)+\mathcal{C}_{h,m}(n,k-1) \left( \frac{\alpha_k}{2} -  \beta_k \right)\\
	&- \mathcal{C}_{h,m}(n,k)\, \gamma \\
	\end{align*}

  If we consider $\gamma=0$, this is just an affine combination. Moreover, we have:
	
	$$
	1 - \alpha_k  \geq 1 -\sigma^2\, 	\displaystyle \frac{h\, 2^{2m}}{2^m \left( \displaystyle \int_L^M \psi_X^{m}d\mu \right)}\geq 0 \quad , \quad 
	\displaystyle  \frac{\alpha_k}{2} - \beta_k  \geq 0 \quad , \quad 
	1-\gamma  \geq 0 \, \cdot\\
	$$
	
  May we suppose that~\mbox{$\sigma^2 \geq r$} and that the following CFL condition
	
	$$ 	\displaystyle \frac{h\, 2^{m}}{\left( \displaystyle\int_L^M \psi_X^{m}d\mu \right)} \varleq	\displaystyle  \frac{1}{\sigma^2}$$
	
	\noindent is satisfied, the combination is then convex, and the scheme is stable for the norm~\mbox{$\|  \cdot  \| _{\infty}$}.\\
	
  For~\mbox{$\gamma \neq 0$}, one has:
	
	\begin{align*}
	(1-\gamma)\min_{0\varleq j \varleq 2^m}   \, \mathcal{C}_{h,m}(n,j) \varleq \mathcal{C}_{h,m}(n+1,k)&\varleq \max_{0\varleq j \varleq 2^m}   \, \mathcal{C}_{h,m}(n,j) - \gamma \, \min_{0\varleq j \varleq 2^m}  \, \mathcal{C}_{h,m}(n,j) \\
	&\varleq \max_{0\varleq j \varleq 2^m}  \, \mathcal{C}_{h,m}(n,j)  - \gamma \, (1-\gamma)^n \, \min_{0\varleq j \varleq 2^m}  \mathcal{C}_{h,m}(0,j)\\
	&\varleq \max_{0\varleq j \varleq 2^m}  \, \mathcal{C}_{h,m}(n,j)\\
	\end{align*}
	
	\noindent and the scheme is~\mbox{$\|  \cdot \| _{\infty}$}-stable under the same conditions.
	
	\vskip 1cm
	
	\paragraph{Convergence}{\ }\\
	
	\vskip 0.25cm
	
	\begin{theorem}{\ }\\
		
  If the above CFL condition holds, the scheme is convergent for the norm $\|  \cdot \| _{2,\infty}$ given by:
		
		$$\|  \left( \mathcal{C}_{h,m}(n,k) \right)_{0\varleq n\varleq N,\, 0\varleq k \varleq 2^m} \| _{2,\infty}= \max_{0\varleq n\varleq N} \left(\sum_{0\varleq k\varleq 2^m}   \mu(f_{\mathcal{W}^k}(\mathcal{M}))\left( \mathcal{C}_{h,m}(n,k)\right)^2)\right)^{\frac{1}{2}}$$
		
		\noindent where $\mu\left(f_{\mathcal{W}^k}(\mathcal{M})\right)$ is the measure of the $f_{\mathcal{W}^k}(\mathcal{M})$.
		
	\end{theorem}
	
	\vskip 1cm

	\begin{proof}{\ }\\
		
 For~\mbox{$0\varleq n \varleq N$} and~\mbox{$0\varleq k\varleq 2^m$}, we set:
		
		$$ w^n_k =  C(n,k)-\mathcal{C}_{h,m}(n,k) $$
		
		\noindent and
		
		\begin{align*}
		\mathcal{BS}_m \, C(n,k)&= -\displaystyle \frac{\sigma^2}{2}\left(\frac{k\,\mathbb{M}}{2^m}\right)^2\left( \frac{C(n,k+1)-2\,C(n,k)+C(n,k-1)}{\left(\frac{\mathbb{M}}{2^m}\right)^2}\right)\\
		& - r\, \left(\frac{k\,\mathbb{M}}{2^m}\right) \left( \frac{C(n,k+1)-C(n,k)}{\left(\frac{\mathbb{M}}{2^m}\right)}\right)+r\,C(n,k)  \, \cdot \\
		\end{align*}
		
We can check that
		\footnotesize
		
		\begin{align*}
		\frac{w^{n+1}_k-w^n_k}{h}-\frac{\sigma^2}{2}\,k^2\,\delta_m\left(w^n_{k+1}-2w^n_k+w^n_{k-1}\right) -r\,k\,\delta_m\left(w^n_{k+1}-w^n_k\right) +r\,\delta_m\,w^n_k&= \varepsilon^{h,m}_{n,k} \quad &0\varleq n \varleq N-1,\, 1\varleq k \varleq 2^m-1 \\
		w^n_0=w^n_{2^m}&=0 \qquad &0\varleq n \varleq N\\
		w^0_k&=0 \qquad &1\varleq k \varleq 2^m-1  \, \cdot 
		\end{align*}
		
		\normalsize
		
  Let us set, for any integer~$n$ in~\mbox{$\left \lbrace 0,\hdots, N \right \rbrace $}:
		
		$$W^n= 
		\left(
		\begin{matrix}
		w^n_1\\
		\vdots \\
		w^n_{2^m-1}
		\end{matrix}
		\right)
		\quad , \quad E^n= 
		\left(
		\begin{matrix}
		\varepsilon^{h,m}_{n,1}\\
		\vdots\\
		\varepsilon^{h,m}_{n,2^m-1}
		\end{matrix}
		\right)  \, \cdot 
		$$
		
  One has:
		
		$$	W^0=0   \quad \text{and} \quad \forall \, n\, \in\, \left \lbrace 0,\hdots, N-1 \right \rbrace :\quad 
		W^{n+1} =A\, W^n+h\, E^n  \, \cdot 
		$$
		
  By induction, this yields, for any integer~$n$ in~\mbox{$\left \lbrace 0,\hdots, N-1 \right \rbrace $}:
		
		$$
		W^{n+1}=A^n \, W^0+h\, \displaystyle \sum_{j=0}^{n} A^j \, E^{n-j}  
		=h\, \displaystyle \sum_{j=0}^{n} A^j \, E^{n-j} 
		$$
		
  Since~$A$ is a symmetric matrix, the \textbf{CFL} stability condition yields, for any integer~$n$ in~\mbox{$\left \lbrace 0,\hdots, N  \right \rbrace $}:
		
		\begin{align*}
		|W^n|&\varleq h\, \displaystyle \left(\sum_{j=0}^{n-1}\|  A \| ^j \right)\left(\max_{0\varleq j \varleq n-1} | E^{j} | \right)\\
		&\varleq h\, n\, \displaystyle \left(\max_{0\varleq j \varleq n-1} | E^{j} | \right)\\
		&\varleq h\, N\, \displaystyle \left(\max_{0\varleq j \varleq n-1} | E^{j} | \right)\\
		&\varleq T\, \displaystyle \left(\max_{0\varleq j \varleq n-1} 
		\left(\sum_{k=1}^{2^m-1} |\varepsilon^{h,m}_{j,k}|^2 \right)^{\frac{1}{2}} \right)  \, \cdot \\
		\end{align*}
		
  By assuming~\mbox{$\displaystyle {\mu_1}\geq \displaystyle\frac{1}{2}$} (the same result holds for $\displaystyle {\mu_1}\varleq \displaystyle \frac{1}{2}$ by changing~${\mu_1}$ into ${\mu_2}$), we deduce then that:
		
		\begin{align*}
		\max_{0\varleq n\varleq N} \left( \sum_{k=1}^{2^m-1} \mu\left(f_{\mathcal{W}^k}(\mathcal{M})\right)  |w^n_k|^2\right)^{\frac{1}{2}}
		&\varleq \max_{1\varleq k \varleq 2^m-1} \left( \mu\left(f_{\mathcal{W}^k}(\mathcal{M})\right)\right)^{\frac{1}{2}}\max_{1\varleq n\varleq N}|W^n|\\
		&\varleq \max_{1\varleq k \varleq 2^m-1}\left( \mu\left(f_{\mathcal{W}^k}(\mathcal{M})\right)\right)^{\frac{1}{2}} T\, \left(\max_{0\varleq n \varleq N-1} \left(\sum_{k=1}^{2^m-1} |\varepsilon^{h,m}_{n,k}|^2 \right)^{1/2} \right)\\
		&\varleq  \displaystyle\max_{1\varleq k \varleq 2^m-1} \left( \mu\left(f_{\mathcal{W}^k}(\mathcal{M})\right) \right)^{\frac{1}{2}} T\, \left((2^m-1)^{\frac{1}{2}}\max_{0\varleq n \varleq N-1,\, 1\varleq k \varleq 2^m-1} |\varepsilon^{h,m}_{n,k}| \right)\\
		&\varleq\sqrt{\left( {\mu_1}\times2\right)^m} T\, \left(\max_{0\varleq n \varleq N-1,\, 1\varleq k \varleq 2^m-1} |\varepsilon^{h,m}_{n,k}| \right)\\
		&= \sqrt{\left( {\mu_1}\times2\right)^m}\left( \mathcal{O}(h) + \mathcal{O}(2^{-m}) \right)\\
		&=\mathcal{O}\left(\left(\sqrt{\frac{{\mu_1}}{2}}\right)^{m}\right)\\
		\end{align*}
		
		\noindent The scheme is then convergent.
	\end{proof}
	
	\vskip 1cm

	\subsection{Self-Similar Pricing}
	
	\hskip 0.5cm In the sequel, we give a numerical simulation of the self-similar pricing, in the case of a call:
	
	\begin{align*}
	\frac{\partial C}{\partial t}(t,S)  & = \mathcal{BS}_{\mu}(C)(t,S) & \forall \, t\,\in \,\left[0,T\right], \,\forall \,S\, \in\, \left[L,M\right[ \\
	C(T,S)&=(S-K)^+ & \forall  \,x \,\in \,\left[L,M\right[\\
	C(t,L)&=0 & \forall  \,t\,\in \, v\left[0,T\right]\\
	C(t,M)&=M - K\exp(-r(T-t))& \forall \,t\,\in \,\left[0,T\right] \\
	\end{align*}
	
	\noindent for a self-similar measure~$\mu$ on~\mbox{$\left[L,M\right]$}. The solutions are generated using the finite difference method, for
	
	$$T=1 \quad , \quad K=150 \quad , \quad \sigma=0.3 \quad , \quad r=0.1 \, \cdot $$
	
	\vskip 1cm
	
	\begin{figure}[!htb]
		
		\minipage{0.7\textwidth}
		\includegraphics[scale=1.5]{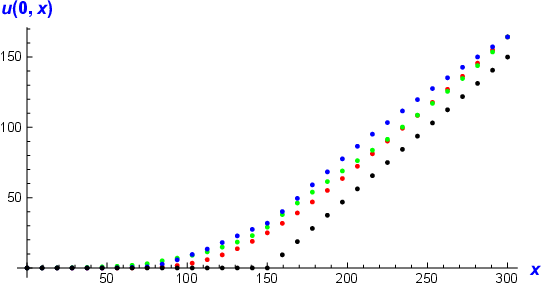}
		\endminipage\hfill
		\minipage{0.2\textwidth}
		\includegraphics[scale=1]{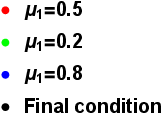}
		\endminipage
		\captionof{figure}{Black-Scholes solution $u(0,x)$ for different values of the weights.}
		\label{fig61}
	\end{figure}
	
	\vskip 1cm

The self-similar Black-Scholes model generates an exotic pricing for different values of $\mu_1$, including the classical one for $\mu_1=\displaystyle\frac{1}{2}$.

	\vskip 1cm
	
	\begin{figure}[!htb]
		
		\minipage{0.3\textwidth}
		\includegraphics[width=\linewidth]{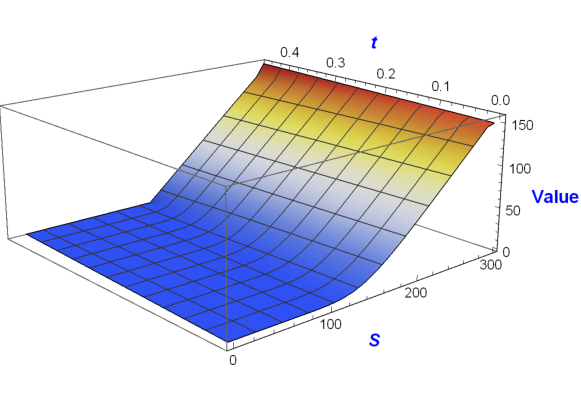}
		\caption*{$\mu_1=\frac{1}{2}$.}
		\endminipage\hfill
		\minipage{0.3\textwidth}
		\includegraphics[width=\linewidth]{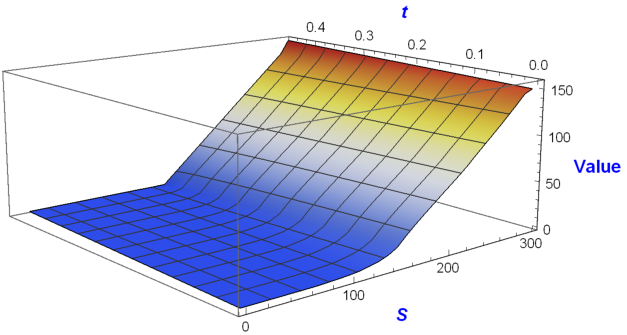}
		\caption*{$\mu_1=0.2$.}
		\endminipage\hfill
		\minipage{0.3\textwidth}
		\includegraphics[width=\linewidth]{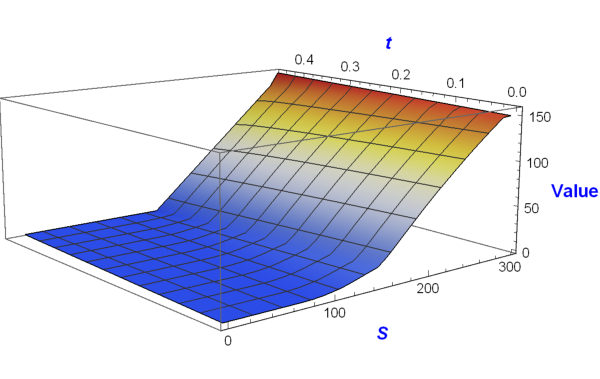}
		\caption*{$\mu_1=0.8$.}
		\endminipage
		\captionof{figure}{The call value for different values of the weights.}
		\label{fig62}
	\end{figure}
	
	\vskip 1cm
	
	\subsection{The Greeks}
	
	\hskip 0.5cm As in~\cite{ClaireBlackScholes}, we recall that, in finance, the sensitivity of a portfolio to changes in parameters values can be measured through what commonly call ``the Greeks", i.e.,
	
	\begin{enumerate}
		
		\item[\emph{i}.] The Delta,~\mbox{$\Delta=\displaystyle \frac{\partial C }{\partial S} \,\in\,[0,1]$}, which enables one to quantify the risk, and is thus the most important Greek. It can also be interpreted as a probability that the option will expire in the money. 
		
		\item[\emph{ii}.] The Gamma,~\mbox{$\Gamma=\displaystyle \frac{\partial^2 C }{\partial  S^2} \geq 0$},  which measures the rate of the acceleration of the option price, with respect to changes in the underlying price.

		\item[\emph{iii}.] The Vega (the name of which comes from the form of the Greek letter $\nu$),~\mbox{$\nu=\displaystyle \frac{\partial  C }{\partial \sigma}$}, which measures the sensitivity to volatility.

		\item[\emph{iv}.] The Theta~\mbox{$\Theta=\displaystyle \frac{\partial  C }{\partial t}$}, which is the time cost of holding an option.

		\item[\emph{ v}.] The rho,~\mbox{$\rho=\displaystyle \frac{\partial  C }{\partial r}$}, which measures the sensitivity to the risk-free interest rate.\\

	\end{enumerate}
	
	The good strategy, for traders, is to have delta-neutral positions at least once a day, and, whenever the opportunity arises, to improve the Gamma and the Vega.
	
	\subsubsection{The Delta}
	
	\begin{figure}[!htb]
		
		\minipage{0.3\textwidth}
		\includegraphics[width=\linewidth]{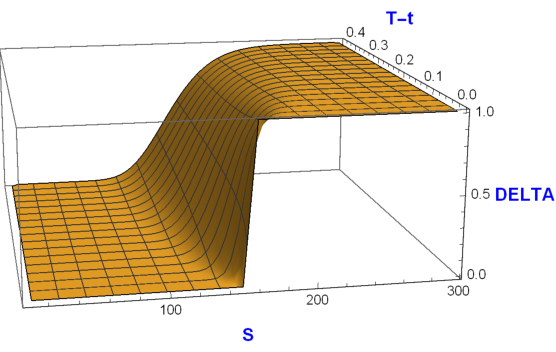}
		\caption*{$\mu_1=\frac{1}{2}$.}
		\endminipage\hfill
		\minipage{0.3\textwidth}
		\includegraphics[width=\linewidth]{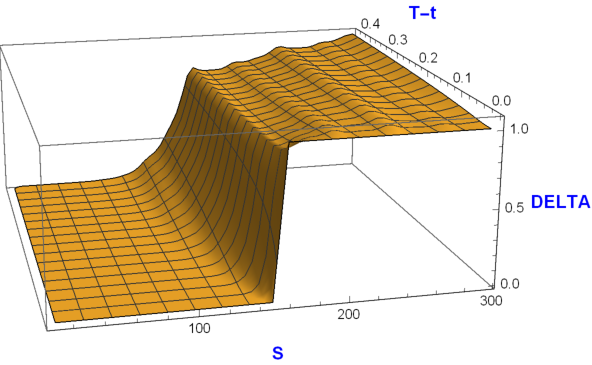}
		\caption*{$\mu_1=\frac{1}{3}$.}
		\endminipage\hfill
		\minipage{0.3\textwidth}
		\includegraphics[width=\linewidth]{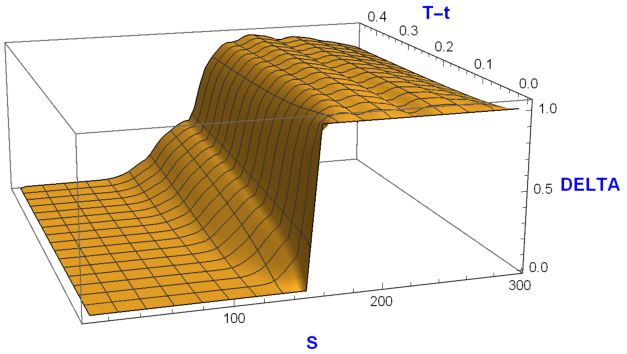}
		\caption*{$\mu_1=\frac{2}{3}$.}
		\endminipage
		\captionof{figure}{The~$\Delta$, for different values of the weights.}
		\label{fig63}
	\end{figure}
	
	\vskip 1cm

	\subsubsection{The Gamma}
	
	\begin{figure}[!htb]
		
		\minipage{0.3\textwidth}
		\includegraphics[width=\linewidth]{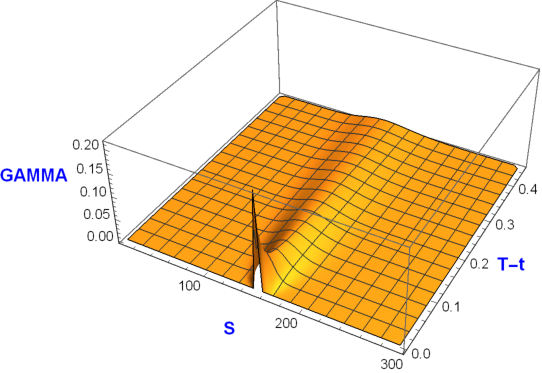}
		\caption*{$\mu_1=\frac{1}{2}$.}
		\endminipage\hfill
		\minipage{0.3\textwidth}
		\includegraphics[width=\linewidth]{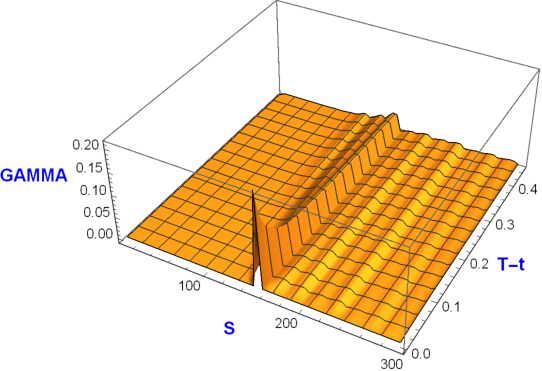}
		\caption*{$\mu_1=\frac{1}{3}$.}
		\endminipage\hfill
		\minipage{0.3\textwidth}
		\includegraphics[width=\linewidth]{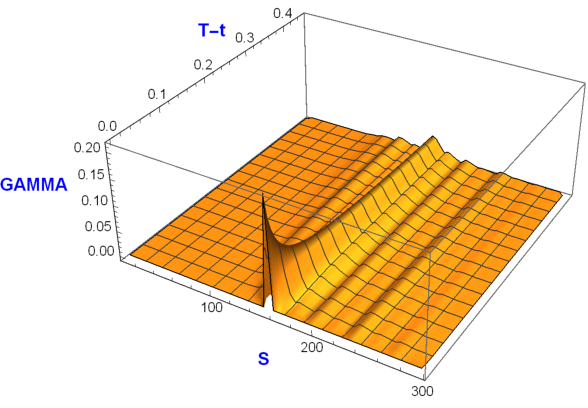}
		\caption*{$\mu_1=\frac{2}{3}$.}
		\endminipage
		\captionof{figure}{The~$\Gamma$, for different values of the weights.}
		\label{fig64}
	\end{figure}
	
	\vskip 1cm

\newpage

	\subsubsection{The Theta}
	
	\begin{figure}[!htb]
		
		\minipage{0.3\textwidth}
		\includegraphics[width=\linewidth]{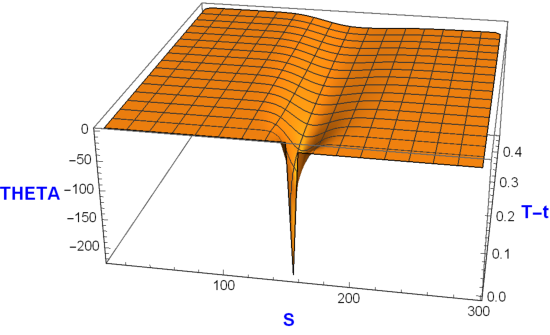}
		\caption*{$\mu_1=\frac{1}{2}$.}
		\endminipage\hfill
		\minipage{0.3\textwidth}
		\includegraphics[width=\linewidth]{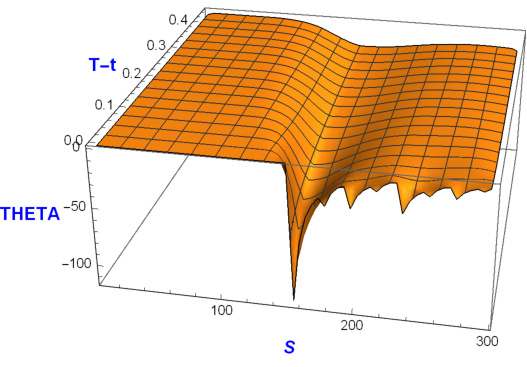}
		\caption*{$\mu_1=\frac{1}{3}$.}
		\endminipage\hfill
		\minipage{0.3\textwidth}
		\includegraphics[width=\linewidth]{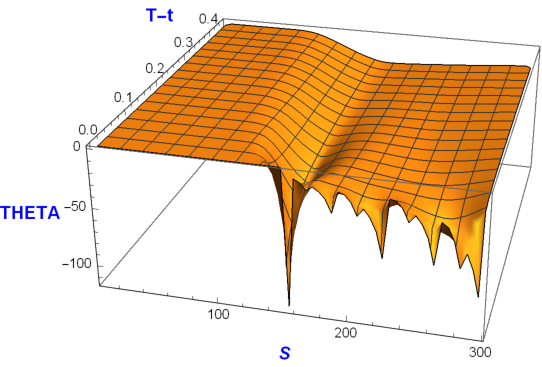}
		\caption*{$\mu_1=\frac{2}{3}$.}
		\endminipage
		\captionof{figure}{The~$\Theta$, for different values of the weights.}
		\label{fig65}
	\end{figure}
	
	\vskip 1cm
	
	\subsubsection{The Vega}
	
	\begin{figure}[!htb]
		
		\minipage{0.3\textwidth}
		\includegraphics[width=\linewidth]{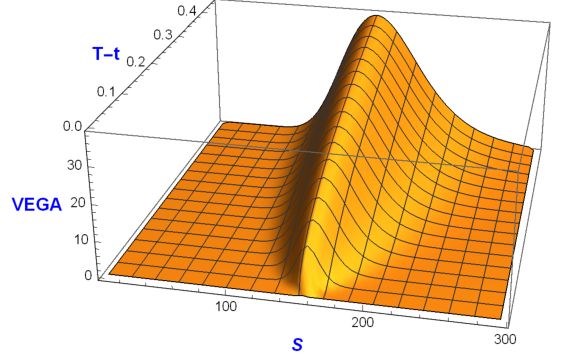}
		\caption*{$\mu_1=\frac{1}{2}$.}
		\endminipage\hfill
		\minipage{0.3\textwidth}
		\includegraphics[width=\linewidth]{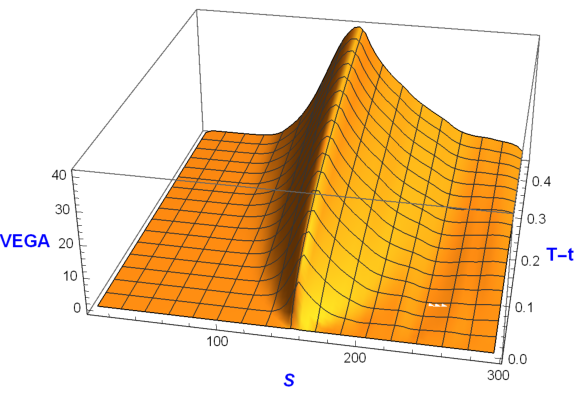}
		\caption*{$\mu_1=\frac{1}{3}$.}
		\endminipage\hfill
		\minipage{0.3\textwidth}
		\includegraphics[width=\linewidth]{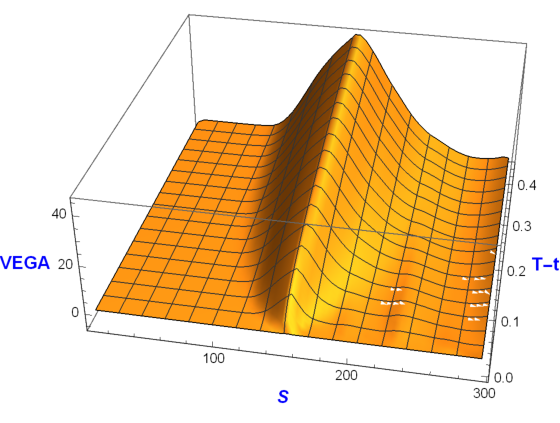}
		\caption*{$\mu_1=\frac{2}{3}$.}
		\endminipage
		\captionof{figure}{The~$\nu$, for different values of the weights.}
		\label{fig66}
	\end{figure}
	
	\vskip 1cm

	\subsubsection{The rho}
	
	\begin{figure}[!htb]
		
		\minipage{0.3\textwidth}
		\includegraphics[width=\linewidth]{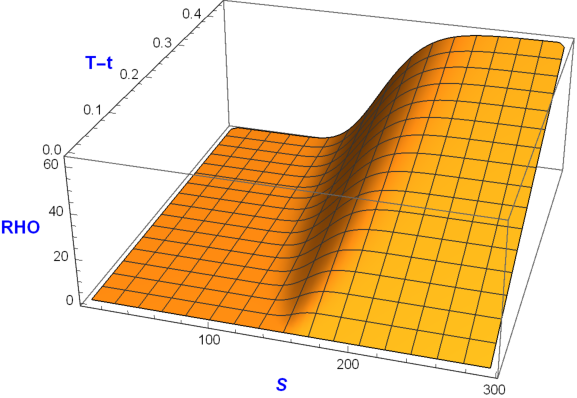}
		\caption*{$\mu_1=\frac{1}{2}$.}
		\endminipage\hfill
		\minipage{0.3\textwidth}
		\includegraphics[width=\linewidth]{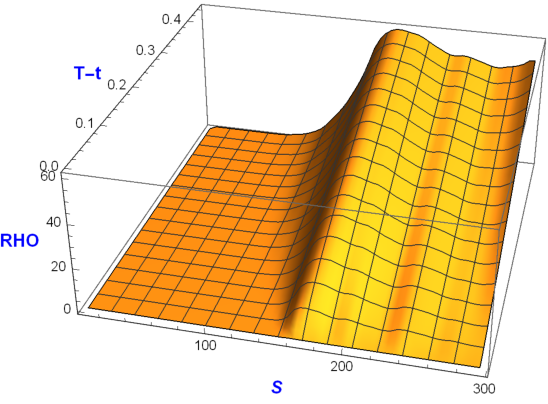}
		\caption*{$\mu_1=\frac{1}{3}$.}
		\endminipage\hfill
		\minipage{0.3\textwidth}
		\includegraphics[width=\linewidth]{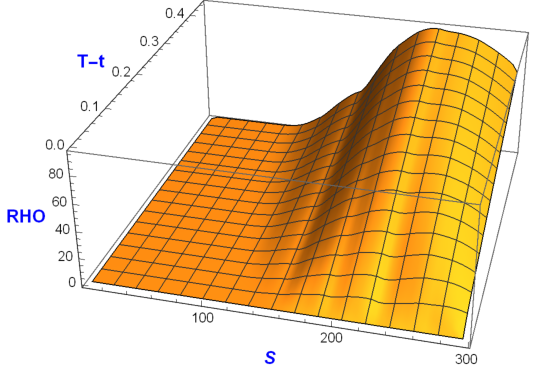}
		\caption*{$\mu_1=\frac{2}{3}$.}
		\endminipage
		\captionof{figure}{The~$\rho$, for different values of the weights.}
		\label{fig67}
	\end{figure}
	
	\vskip 1cm
	
	\subsection{Discussion }
	
	 \hskip 0.5cm Let us make a few remarks about the behavior of the solution with respect to the weight $\mu$ at $t=0$:\\

	\begin{enumerate}

		\item[\emph{i}.]  For~$\displaystyle\mu_1=\frac{1}{3}$: the premium is greater than that of the classical model under the strike and smaller above. The Greeks value shows a drastic increase in the strike neighbor, and self-similarity clearly affects \emph{in the money} region. The Theta shows a slower premium expected decrease.\\
		
		\item[\emph{ii}.]  For~$\displaystyle\mu_1=\frac{2}{3}$: the premium is everywhere greater than that of the classical model. The Greeks value increases progressively in the strike neighbor, and self-similarity affects \emph{in the money} region. The Theta indicates a slower premium expected decrease \emph{in the money} and a greater decrease \emph{deep in the money}.\\
		
	\end{enumerate}
	
 The dynamic generated by the self-similar Black-Scholes model is exotic and enables the emergence of non-standard behaviors, the parameter $\displaystyle\mu_1 $ can capture the behavior of non confident investors under \textbf{uncertainty} and other factors influencing their perception of future.\\
	
 The self-similar Black-Scholes equation can be understood as a diffusion equation with a time change through self-similar probability $\mu$, where the cumulative distribution function satisfies for $x\in\left]0,1\right[$, $\left[0,x\right]=f_{\mathcal{W}}(\left[0,1\right])$ for some word $\mathcal{W}$,
	
\begin{align*}
\mu\left[0,x\right]&=\Pi_{i\in \mathcal{W}} \, \mu_i\\
\end{align*}
	
\noindent depending on the address (path) of $x$. According to this remark, we can create more exotic behavior by using a self-similar measure $\mu$ with many weights or enable the weights to change over time.\\

\vskip 1cm

	\bibliographystyle{alpha}
	\bibliography{BibliographieClaire}

\begin{thebibliography}{LMM05}

\bibitem[AP05]{Achdou2005}
Y.~Achdou and O.~Pironneau.
\newblock {\em Computational Methods for Option Pricing}.
\newblock Society for Industrial and Applied Mathematics, 2005.

\bibitem[Dav17]{ClaireBlackScholes}
C.~David.
\newblock Control of the {B}lack-{S}choles equation.
\newblock {\em Comput. Math. Appl.}, 73(7):1566--1575, 2017.

\bibitem[For96]{Fortune1996}
P.~Fortune.
\newblock Anomalies in option pricing: The black-scholes model revisited.
\newblock {\em New England Economic Review}, March/April:17--40, 1996.

\bibitem[HLN06]{Hu2006}
J.~Hu, K.-S. Lau, and S.-M. Ngai.
\newblock Laplace operators related to self-similar measures on {$\R^d$}.
\newblock {\em Journal of Functional Analysis}, 239:542–565, 2006.

\bibitem[Kig01]{Kigami2001}
J.~Kigami.
\newblock {\em Analysis on {F}ractals}.
\newblock Cambridge University Press, 2001.

\bibitem[Kil94]{Kilpelainen1994}
T.~Kilpel\"{a}inen.
\newblock Weighted sobolev spaces and capacity.
\newblock {\em Annales Academiae Scientiarum Fennicae. Series A I.
  Mathematica}, 19(1):95--113, 1994.

\bibitem[KN01]{Kangro2001}
R.~Kangro and R.~Nicolaides.
\newblock Far field boundary conditions for black-scholes equations.
\newblock {\em SIAM Journal on Numerical Analysis}, 38(4):1357--1368, 2001.

\bibitem[LM68]{Lions1968}
J.-L. Lions and E.~Magenes.
\newblock {\em Problèmes aux limites non homogènes et applications}.
\newblock Dunod, 1968.

\bibitem[LMM05]{Martin2005}
G.~C. Lim, G.~M. Martin, and V.~L. Martin.
\newblock Parametric pricing of higher order moments in s\&p500 options.
\newblock {\em Journal of Applied Econometrics}, 20(3):377--404, 2005.

\bibitem[Man97]{Mandelbrot1997}
B.~B. Mandelbrot.
\newblock {\em Fractals and Scaling in Finance}.
\newblock Springer New York, 1997.

\bibitem[MR92]{Ma1992}
Z.-M. Ma and M.~R\"{o}ckner.
\newblock {\em Introduction to the {T}heory of ({N}on-{S}ymmetric) {D}irichlet
  {F}orms}.
\newblock Springer-Verlag, 1992.

\bibitem[RD19]{RianeDavidFDMS}
N.~Riane and C.~David.
\newblock The finite difference method for the heat equation on sierpiński
  simplices.
\newblock {\em International Journal of Computer Mathematics},
  96(7):1477--1501, 2019.

\bibitem[RD21]{RianeInv}
N.~Riane and C.~David.
\newblock An inverse black–scholes problem.
\newblock {\em Optimization and Engineering}, 22:2183–2204, 2021.

\bibitem[Str06]{StrichartzLivre2006}
R.~S. Strichartz.
\newblock {\em Differential {E}quations on {F}ractals. {A} {T}utorial}.
\newblock Princeton University Press, 2006.

\bibitem[Wlo87]{Wolka1987}
J.~Wloka.
\newblock {\em Partial differential equations}.
\newblock Cambridge University Press, 1987.

\bibitem[Zei90]{Zeidler1990}
E.~Zeidler.
\newblock {\em Nonlinear {F}unctional {A}nalysis and its {A}pplications. II/A.:
  {L}inear {M}onotone {O}perators}.
\newblock Springer, 1990.

\end{thebibliography}
\end{document}